\documentclass[11pt]{article}
\usepackage[letterpaper,margin=1in]{geometry}
\usepackage[T1]{fontenc}
\usepackage{lmodern,microtype}
\usepackage{amsmath,amssymb,amsthm,mathtools}
\usepackage{enumitem,booktabs,array}
\usepackage{cite}
\usepackage{placeins}
\usepackage{xurl}
\usepackage[dvipsnames]{xcolor}
\usepackage{tikz}
\usetikzlibrary{arrows.meta}
\usepackage[colorlinks=true,linkcolor=MidnightBlue,citecolor=MidnightBlue,urlcolor=MidnightBlue]{hyperref}
\usepackage[nameinlink,noabbrev]{cleveref}
\hypersetup{pdftitle={A Common Improved Bound for Asymmetric TSP and Path TSP},pdfauthor={Mohammad Hajiaghayi and Changki Yun}}
\newtheorem{theorem}{Theorem}[section]
\newtheorem{lemma}[theorem]{Lemma}
\newtheorem{proposition}[theorem]{Proposition}
\newtheorem{corollary}[theorem]{Corollary}

\theoremstyle{remark}
\newtheorem{remark}[theorem]{Remark}
\DeclareMathOperator{\diver}{div}

\DeclareMathOperator{\LP}{LP}
\DeclareMathOperator{\cov}{cov}

\setlist{itemsep=3pt,topsep=5pt}
\numberwithin{equation}{section}
\title{A Common Improved Bound for\\Asymmetric TSP and Path TSP}
\author{Mohammad Hajiaghayi \qquad Changki Yun\\
University of Maryland, College Park\\
\texttt{hajiagha@umd.edu} \qquad \texttt{ckyun@umd.edu}}
\date{}
\definecolor{AtspBlue}{RGB}{27,83,130}
\definecolor{AtspOchre}{RGB}{166,82,24}
\definecolor{AtspGreen}{RGB}{37,112,82}
\usetikzlibrary{decorations.markings}
\tikzset{
  atsp vertex/.style={circle,fill=black,inner sep=1.5pt},
  atsp unvisited/.style={circle,draw=black!55,fill=white,inner sep=1.6pt},
  atsp region/.style={draw=black!45,line width=.55pt,fill=black!2,rounded corners=2pt},
  atsp blue/.style={draw=AtspBlue,line width=1.05pt,-{Stealth[length=2.1mm]}},
  atsp orange/.style={draw=AtspOchre,line width=1.05pt,dash pattern=on 4pt off 1.5pt on 1pt off 1.5pt,-{Stealth[length=2.1mm]}},
  atsp green/.style={draw=AtspGreen,line width=1.5pt,-{Stealth[length=2.2mm]}},
  atsp green cycle/.style={draw=AtspGreen,line width=1.5pt,
    postaction={decorate},decoration={markings,
      mark=at position .56 with {\arrow{Stealth[length=1.8mm]}}}},
  atsp connector/.style={draw=black!75,line width=.85pt,densely dashed,-{Stealth[length=1.9mm]}},
  atsp operation/.style={draw=black!60,line width=.8pt,-{Stealth[length=2mm]}}
}

\newcommand{\AtspLiftedCycles}[1]{%
  \draw[atsp region] (-2.3,-1.75) rectangle (2.3,1.75);
  \node[anchor=north west,inner sep=3pt] at (-2.3,1.75) {$L$};
  \coordinate (b1) at (-1.1,2.45);
  \coordinate (b2) at (0,3.0);
  \coordinate (b3) at (1.1,2.45);
  \coordinate (c1) at (-1.1,-2.45);
  \coordinate (c2) at (0,-3.0);
  \coordinate (c3) at (1.1,-2.45);
  \coordinate (p1) at (-1.95,.9);
  \coordinate (p2) at (-1.2,1.25);
  \coordinate (p3) at (-.55,.85);
  \coordinate (p4) at (.2,1.2);
  \coordinate (p5) at (1.0,.9);
  \coordinate (p6) at (1.95,1.2);
  \coordinate (q1) at (-1.95,-1.15);
  \coordinate (q2) at (-1.15,-.85);
  \coordinate (q3) at (-.45,-1.25);
  \coordinate (q4) at (.35,-.9);
  \coordinate (q5) at (1.1,-1.15);
  \coordinate (q6) at (1.95,-.75);
  \coordinate (z1) at (-.4,.1);
  \coordinate (z2) at (.6,-.2);
  \foreach \a/\b in {b2/b1,b1/p1,p1/p2,p2/p3,p3/p4,p4/p5,p5/p6,p6/b3,b3/b2}
    \draw[atsp blue] (\a) -- (\b);
  \foreach \a/\b in {c2/c1,c1/q1,q1/q2,q2/q3,q3/q4,q4/q5,q5/q6,q6/c3,c3/c2}
    \draw[atsp orange] (\a) -- (\b);
  \node[text=AtspBlue,anchor=west] at (.9,2.9) {$B_L$};
  \ifnum#1=1
    \foreach \a/\b in {p3/z1,z1/z2,z2/q6,q1/p3}
      \draw[atsp green] (\a) -- (\b);
    \foreach \a/\b in {q6/q5,q5/q4,q4/q3,q3/q2,q2/q1}
      \draw[atsp green] (\a) to[bend left=55] (\b);
    \foreach \a in {z1,z2} \node[atsp vertex] at (\a) {};
  \else
    \foreach \a in {z1,z2} \node[atsp unvisited] at (\a) {};
  \fi
  \foreach \a in {b1,b2,b3,c1,c2,c3,p1,p2,p3,p4,p5,p6,q1,q2,q3,q4,q5,q6}
    \node[atsp vertex] at (\a) {};
}

\newcommand{\LiftingFigure}{%
\begin{figure}[!htb]
\centering
\begin{tikzpicture}[scale=1.07,x=1.05cm,y=1.05cm,font=\small,
  every node/.style={inner sep=1.5pt}]
  \begin{scope}[xshift=1.4cm]
    \coordinate (b1) at (-1.1,2.45);
    \coordinate (b2) at (0,3.0);
    \coordinate (b3) at (1.1,2.45);
    \coordinate (c1) at (-1.1,-2.45);
    \coordinate (c2) at (0,-3.0);
    \coordinate (c3) at (1.1,-2.45);
    \node[circle,draw=black!65,fill=black!8,minimum size=8mm] (L) at (0,0) {$v_L$};
    \draw[atsp blue] (b2) -- (b1);
    \draw[atsp blue] (b1) .. controls (-1.2,1.6) and (-.8,.7) .. (L.north west);
    \draw[atsp blue] (L.north east) .. controls (.8,.7) and (1.2,1.6) .. (b3);
    \draw[atsp blue] (b3) -- (b2);
    \draw[atsp orange] (c2) -- (c1);
    \draw[atsp orange] (c1) .. controls (-1.2,-1.6) and (-.8,-.7) .. (L.south west);
    \draw[atsp orange] (L.south east) .. controls (.8,-.7) and (1.2,-1.6) .. (c3);
    \draw[atsp orange] (c3) -- (c2);
    \foreach \a in {b1,b2,b3,c1,c2,c3} \node[atsp vertex] at (\a) {};
  \end{scope}
  \draw[atsp operation] (2.8,0) -- (3.7,0);
  \node[above=5pt] at (3.25,0) {lift};
  \begin{scope}[xshift=6.5cm]
    \AtspLiftedCycles{0}
  \end{scope}
  \draw[atsp operation] (8.7,0) -- (9.3,0);
  \node[above=5pt,text=AtspGreen] at (9.0,0) {$+E_L$};
  \begin{scope}[xshift=12.4cm]
    \AtspLiftedCycles{1}
  \end{scope}
\end{tikzpicture}
\caption{Lifting can disconnect circuits; internal Eulerian augmentation restores connectivity.}
\label{fig:lifting}
\end{figure}%
}

\newcommand{\PathWalksFigure}{%
\begin{figure}[tb]
\centering
\begin{tikzpicture}[x=1cm,y=1cm,font=\small,
  every node/.style={inner sep=1.5pt}]
  \node[anchor=west] at (0,5.2) {$Q_{\mathrm{keep}}$};
  \foreach \x/\j/\parity in {.9/i/1,8.2/{i+1}/0} {
    \begin{scope}[xshift=\x cm]
      \draw[atsp region] (0,1.2) rectangle (6.3,4.7);
      \node[anchor=west] at (.2,4.45) {$W_{\j}$};
      \coordinate (a0) at (.4,3.7);
      \coordinate (a1) at (1.5,4.0);
      \coordinate (a2) at (2.5,3.55);
      \coordinate (a3) at (3.6,4.0);
      \coordinate (a4) at (4.7,3.65);
      \coordinate (a5) at (5.9,3.7);
      \coordinate (b0) at (.4,1.8);
      \coordinate (b1) at (1.5,2.15);
      \coordinate (b2) at (2.5,1.6);
      \coordinate (b3) at (3.6,2.05);
      \coordinate (b4) at (4.7,1.65);
      \coordinate (b5) at (5.9,1.8);
      \foreach \u/\v in {a0/a1,a1/a2,a2/a3,a3/a4,a4/a5}
        \draw[atsp blue] (\u) -- (\v);
      \foreach \u/\v in {b0/b1,b1/b2,b2/b3,b3/b4,b4/b5}
        \draw[atsp orange] (\u) -- (\v);
      \node[text=AtspBlue] at (3.15,4.35) {$P_1$};
      \node[text=AtspOchre] at (3.15,1.4) {$P_2$};
      \ifnum\parity=1
        \coordinate (k1) at (4.7,2.85);
        \coordinate (k2) at (3.4,2.8);
        \coordinate (k3) at (2.0,2.7);
        \draw[atsp connector] (a5) -- (k1);
        \draw[atsp connector] (k3) -- (b0);
      \else
        \coordinate (k1) at (4.7,2.5);
        \coordinate (k2) at (3.4,2.8);
        \coordinate (k3) at (2.0,2.7);
        \draw[atsp connector] (b5) -- (k1);
        \draw[atsp connector] (k3) -- (a0);
      \fi
      \draw[atsp connector] (k1) -- (k2);
      \draw[atsp connector] (k2) -- (k3);
      \foreach \p in {a0,a1,a2,a3,a4,a5,b0,b1,b2,b3,b4,b5,k1,k2,k3}
        \node[atsp vertex] at (\p) {};
    \end{scope}
  }
  \draw[atsp blue] (.1,3.7) -- (1.3,3.7);
  \draw[atsp orange] (6.8,1.8) -- (8.6,1.8);
  \draw[atsp blue] (14.1,3.7) -- (15.3,3.7);
  \node at (-.12,3.7) {$\cdots$};
  \node at (15.52,3.7) {$\cdots$};

  \node[anchor=west] at (0,.6) {$Q_{\mathrm{max}}$};
  \foreach \x/\j in {.9/i,8.2/{i+1}} {
    \begin{scope}[xshift=\x cm]
      \draw[atsp region] (0,-3.4) rectangle (6.3,.1);
      \node[anchor=west] at (.2,-.15) {$W_{\j}$};
      \coordinate (s) at (.4,-.9);
      \coordinate (u) at (1.3,-2.3);
      \coordinate (p1) at (2.2,-1.8);
      \coordinate (p2) at (3.1,-2.45);
      \coordinate (p3) at (4.0,-2.0);
      \coordinate (v) at (5.0,-2.55);
      \coordinate (t) at (5.9,-.9);
      \coordinate (k1) at (.8,-1.6);
      \coordinate (k2) at (5.5,-1.8);
      \foreach \a/\b in {s/k1,k1/u,v/k2,k2/t}
        \draw[atsp connector] (\a) -- (\b);
      \foreach \a/\b in {u/p1,p1/p2,p2/p3,p3/v}
        \draw[atsp green] (\a) -- (\b);
      \foreach \p in {s,u,p1,p2,p3,v,t,k1,k2}
        \node[atsp vertex] at (\p) {};
      \node[text=AtspGreen] at (3.15,-1.15) {$P^{W_{\j}}_{u_{\j}v_{\j}}$};
    \end{scope}
  }
  \draw[atsp blue] (.1,-.9) -- (1.3,-.9);
  \draw[atsp blue] (6.8,-.9) -- (8.6,-.9);
  \draw[atsp blue] (14.1,-.9) -- (15.3,-.9);
  \node at (-.12,-.9) {$\cdots$};
  \node at (15.52,-.9) {$\cdots$};
\end{tikzpicture}
\caption{The two initial walks ($i$ odd). Dashed segments are added nice paths.}
\label{fig:path-walks}
\end{figure}%
}

\begin{document}
\maketitle
\begin{abstract}
We give polynomial-time $(R_*+\varepsilon)$-approximation algorithms for
asymmetric TSP and asymmetric path TSP, for every fixed $\varepsilon>0$, where
$R_*=6+\sqrt2+\sqrt{15+6\sqrt2}\approx12.2604$.
Both guarantees hold relative to the corresponding subtour relaxations,
whose integrality ratios are therefore at most $R_*$.
This improves the previous guarantees of below $14.7$ for asymmetric TSP
and $15$ for asymmetric path TSP [Vygen, 2026].
Our algorithms strengthen subtour covers and recursive augmentation in
the laminar framework [Svensson, Tarnawski, V\'egh, JACM'20;
Traub, Vygen, 2025]. The recursive analysis retains the paths through
contracted regions and accounts jointly for their cost and the dual mass
they visit. We also improve the reduction from asymmetric path TSP to
vertebrate pairs [Traub, Vygen, 2025] by comparing two initial walks with
complementary completion bounds. Applied to Vygen's existing vertebrate-pair
algorithm, this reduction alone improves the path factor from $15$ to
below $14.7$, matching his asymmetric TSP guarantee.
\end{abstract}

\section{Introduction}
The asymmetric traveling salesman problem (ATSP) asks for a minimum-cost tour
visiting every vertex of a directed graph with nonnegative arc costs. In
asymmetric path TSP, the walk must start at a given vertex $s$ and end
at a given distinct vertex $t$. We allow repeated vertices and arcs; on a
complete directed metric, shortcutting gives a Hamiltonian tour or an
$(s,t)$-path.

The first constant-factor approximation for ATSP was obtained by Svensson,
Tarnawski, and V\'egh~\cite{STV20}. Their laminar framework was subsequently
improved by Traub and Vygen~\cite{TV22,TV25} and by
Vygen~\cite{Vyg26}, who gives a tour factor below $14.7$ and an $(s,t)$-path
factor of $15$. We improve both guarantees to a common factor. Let
$\LP(G,c)$ and $\LP_{s,t}(G,c)$ denote the optima of the standard tour and
path subtour relaxations, defined in Section~\ref{sec:preliminaries}, and set
\[
 R_* = 6+\sqrt2+\sqrt{15+6\sqrt2}\approx12.2604.
\]
\begin{theorem}[Approximation guarantees]
\label{thm:main}
For every fixed $\varepsilon>0$, there are polynomial-time algorithms that, on
nonnegative rational arc costs, return a spanning tour of cost at most
$(R_*+\varepsilon)\LP(G,c)$ and a spanning $s$--$t$ walk of cost at most
$(R_*+\varepsilon)\LP_{s,t}(G,c)$, respectively. The ATSP input is strongly
connected; the path input has a feasible path relaxation. The integrality
ratio of each relaxation is at most $R_*$.
\end{theorem}

Our tour improvement builds on Svensson's algorithm~\cite{Sve15}, as
formulated for \emph{vertebrate pairs} by Traub and
Vygen~\cite[Algorithm~7.19]{TV25}. Such a pair consists of a strongly laminar instance and a connected
Eulerian backbone $B$ meeting every nonsingleton laminar set. Following
Traub and Vygen~\cite[Definition~7.7]{TV25}, a $(\kappa,\eta)$-algorithm
adds arcs $H$ so that $B\cup H$ is a spanning tour and
\[
 c(H)\le\kappa\LP+\eta\sum_{v\notin V(B)}2y_v,
\]
where $y_v$ is the singleton dual weight at $v$. Thus $\kappa$ charges
the LP value, and $\eta$ charges the singleton weight missed by the backbone.
Our strengthened subtour cover gives a $(2,\eta)$-algorithm for every
$\eta>3+2\sqrt2$, improving Vygen's $\eta>4+2\sqrt2$ guarantee
\cite[Theorem~3]{Vyg26}. We further improve the recursive reduction to
vertebrate pairs by retaining the paths chosen through contracted regions.
The dual weight visited by these paths reduces the cost of completing those
regions, yielding the tour bound in Theorem~\ref{thm:main}.

Our second result strengthens the reduction from asymmetric path TSP to
vertebrate pairs. It applies to any algorithm with the standard guarantee
above, independently of our subtour-cover construction.

\begin{theorem}[Transfer from vertebrate pairs]
\label{thm:intro-transfer}
Let $\kappa\ge1$ and $\eta\ge0$ be fixed rational constants.
A polynomial-time $(\kappa,\eta)$-algorithm for vertebrate pairs gives
polynomial-time LP-relative approximation algorithms with factors
\[
 \begin{aligned}
  \text{ATSP:}\qquad &3\kappa+\eta+2,\\
  \text{asymmetric path TSP:}\qquad
  &\max\left\{3\kappa+\eta+2,\;
       4\kappa+5-\frac{4\kappa+4}{2\kappa+\eta+2}\right\}.
 \end{aligned}
\]
\end{theorem}

The tour factor is the established reduction to vertebrate
pairs~\cite[Theorem~7.12]{TV25}; our
contribution is the smaller path factor. The proof also accommodates the
refined recursive reduction of Traub and Vygen~\cite[Exercises~7.5--7.7]{TV25},
giving a path guarantee that matches Vygen's refined tour bound
(Corollary~\ref{cor:path-vygen-refined}). Finally, the same path construction
uses our stronger recursive estimate to obtain the path bound in
Theorem~\ref{thm:main}. This last application uses the retained-path analysis
in addition to the $(2,\eta)$-guarantee.

\paragraph{Relation to previous work.}
Table~\ref{tab:history} summarizes the recent approximation guarantees.
Earlier tour-to-path reductions relate the two problems in several ways.
Feige and Singh~\cite[Theorem~2]{FS07} give a general directed reduction with
a multiplicative loss of $2+\varepsilon$. At the level of the subtour LP,
K\"ohne, Traub, and Vygen~\cite{KTV20} prove that a tour integrality ratio
of $\rho$ implies a path integrality ratio of at most $4\rho-3$.
For the symmetric problem, Traub, Vygen, and Zenklusen~\cite{TVZ22} preserve
the approximation ratio up to an arbitrarily small additive error.

More closely related to our work, Traub and Vygen~\cite[Theorem~9.23 and
Corollary~9.24]{TV25} already obtained a common $17+\varepsilon$ guarantee
for asymmetric tours and paths. Their reduction uses the guarantee for
augmenting a partial solution, rather than only a complete-tour approximation
factor. Its application to Vygen's algorithm gives a path factor of $15$,
so it does not carry his improved tour factor over to paths.

\begin{table}[tb]
\centering
\small
\renewcommand{\arraystretch}{1.15}
\begin{tabular}{@{}p{0.56\linewidth}rr@{}}
\toprule
Work & ATSP & Asymmetric path TSP\\
\midrule
Thin-tree rounding~\cite{AGMOS10}
 & $O(\log n/\log\log n)$ & ---\\
Svensson, Tarnawski, and V\'egh~\cite{STV20};
K\"ohne, Traub, and Vygen~\cite{KTV20}
 & $506$ & $2021$\\
Traub and Vygen~\cite{TV22} & $22+\varepsilon$ & ---\\
Traub and Vygen~\cite{TV25} & $17+\varepsilon$ & $17+\varepsilon$\\
Vygen~\cite{Vyg26} & $<14.7$ & $15$\\
\textbf{This paper} & $R_*+\varepsilon$ & $R_*+\varepsilon$\\
\bottomrule
\end{tabular}
\caption{Selected LP-relative approximation guarantees, for fixed
$\varepsilon>0$. The first algorithm is randomized; the others are deterministic.
The first path entry uses the reduction of K\"ohne, Traub, and Vygen, as stated
in~\cite[Corollary~11.4]{STV20}. A dash indicates that no path bound is listed.}
\label{tab:history}
\end{table}

Thin-tree rounding provides another approach to ATSP~\cite{AGMOS10}.
OpenAI's recent preprint~\cite{OAI26} gives a deterministic polynomial-time
construction of strong thin trees with an unspecified universal constant.
This provides another route to constant-factor approximation. Our explicit
bounds refine the recursive laminar framework.

\paragraph{Organization.}
Section~\ref{sec:overview} gives the technical overview;
Section~\ref{sec:preliminaries} fixes notation.
Sections~\ref{sec:geometry}--\ref{sec:recursion} prove the boundary-closure,
subtour-cover, and recursive-augmentation results.
Section~\ref{sec:guarantees} proves the transfer and approximation guarantees.
Appendix~\ref{app:hubs} treats concentrated boundary ports.

\section{Technical overview}
\label{sec:overview}
The tour improvement comes from accounting for a route through a contracted
region together with the cost of completing that region. We first recall
why the existing algorithms use this recursive representation. We then
explain how retaining the actual routes improves the recursive analysis,
how the stronger subtour cover pays for the connections, and how the
resulting augmentation guarantee transfers to $(s,t)$-paths.

\subsection{The recursive framework}
The subtour LP supplies a fractional solution and a lower bound against
which to charge the algorithm. To make this lower bound useful locally,
Svensson, Tarnawski, and V\'egh~\cite{STV20} reduce ATSP to instances whose
arc costs are represented by weighted cuts. The sets defining these cuts
form a \emph{laminar family}: any two are nested or disjoint. The reduction
preserves the LP-relative approximation factor
\cite[Theorem~7.4]{TV25}. We refer to it as the \emph{laminar reduction}.
After this reduction, an arc costs the sum of the weights of the cuts it
crosses. These cuts are tight for the supplied fractional solution, so each
contributes twice its weight to the LP value. We call this contribution its
\emph{dual mass}.
The nested sets can now serve as regions for a recursive algorithm, with
an explicit share of the lower bound assigned to each region.

The recursion starts with a connected Eulerian multigraph, called a
\emph{partial backbone}, and extends it until it spans the graph.
Within a region $W$, it contracts the inclusion-maximal proper
positive-weight laminar sets missed by the partial backbone. The contracted
graph is a simpler instance in which these missed regions are represented
by vertices. A solution on this quotient is lifted by replacing every visit
to a contracted vertex by a path between its original entry and exit.
The contracted regions are then completed recursively. This architecture
is inherited from Svensson, Tarnawski, and V\'egh~\cite[Section~9.2]{STV20}
and Traub and Vygen~\cite[Algorithm~7.8 and Lemma~7.13]{TV25}.

The quotient is where Svensson's algorithm is used: the image of the
backbone meets every remaining nonsingleton laminar set, giving a
vertebrate pair. On such an instance, the algorithm maintains a local
Eulerian solution outside the backbone. Here \emph{local} means that an
arc crosses no nonsingleton laminar cut. If $y_v$ denotes the weight of the
singleton cut at $v$, a local arc $uv$ has cost $y_u+y_v$.
A simple local circuit $C$ therefore costs $\sum_{v\in V(C)}2y_v$;
for a general local Eulerian multiset, the same accounting counts each visit.
This identity lets the algorithm allocate vertex budgets and charge circuits
to the components they meet. A subtour cover supplies connections between
the current components and to the backbone. Its components meeting the
backbone may use nonlocal arcs. Vygen~\cite[Sections~2--4]{Vyg26} improves the
initialization and charging analysis within this framework.

\subsection{Retaining passage costs and recursive progress}
In a recursive call, a quotient vertex represents an entire laminar region.
Its entry and exit determine which internal path must be restored when the
solution is lifted. For each endpoint pair, the reduction fixes one of the
\emph{nice paths} of Traub and Vygen~\cite[Lemma~7.5]{TV25}: it lies in the
smallest laminar set containing its endpoints (allowing the whole graph)
and crosses each laminar
boundary at most once in each direction. The basic recursive reduction
assigns a fixed charge to the quotient vertex large enough to pay for every
such path. Our extension of Svensson's algorithm retains the actual entry,
exit, and path of each selected passage instead.

This information links the cost of lifting to progress within the region.
The nice-path identity~\cite[Lemma~7.6]{TV25} equates the path cost, together
with its entry and exit contributions, to the dual mass of the laminar sets
it visits. We call this visited mass its \emph{coverage}. A passage of small
coverage is cheap but leaves more mass for recursive completion; a passage
of large coverage is more expensive but can reduce the completion cost.
Our extension of the initialization and charging analysis bounds the total
cost of the retained passages while keeping this coverage available to the
recursive analysis.

Coverage must belong to a connected object from which recursion can
continue. Quotient connectivity alone does not provide such an object:
two circuits meeting at one contracted vertex may lift to disjoint closed
walks. For each contracted region $L$, we select a retained passage of
maximum coverage. When processing $L$, the whole current connected Eulerian
component containing that passage is a possible backbone $B_L$; it may
extend outside $L$. Its coverage is certified before the recursive call.
Augmenting $B_L$ with arcs inside $L$ to visit every vertex of $L$ then joins
all other lifted components meeting $L$.
Figure~\ref{fig:lifting} illustrates the loss of connectivity under lifting
and its restoration by an internal augmentation.

\LiftingFigure

The recursive algorithm compares completing this inherited component with
starting from a fresh partial backbone, as in the refined reduction of
Traub and Vygen~\cite[Exercises~7.5--7.6]{TV25}. Retaining passage-dependent
costs improves the accounting of this choice. The resulting augmentation
bound decreases with the dual mass already visited by the inherited
backbone (Theorem~\ref{thm:augment}). Svensson's algorithm and the recursive
completion have distinct roles: the former joins components in the quotient,
whereas the latter completes contracted regions in the original graph.

\subsection{Paying for subtour-cover repairs}
To apply the subtour-cover algorithm inside a recursive region, we first
need a fractional comparison circulation there. Restricting the original
flow to the region can destroy balance. We choose and price internal return
paths to restore it, while retaining the attachment information needed when
the quotient solution is lifted (Section~\ref{sec:geometry}). This construction
also accounts for the fresh backbone choices used by the recursion.

The subtour cover used by Vygen rounds a modified fractional circulation
and then repairs the entry--exit pairing chosen by rounding. Its cost bound
pays for these repairs through an additional singleton-mass term
\cite[Theorem~8.24]{TV25}. We include the repair prices in the rounding
objective and obtain
\[
 c(F)\le 2c^{\mathsf T}x,
\]
where $F$ is the cover and $x$ is the supplied fractional comparison
circulation. The difficulty is that rounding may choose a different pairing
from the fractional one. We price all possible repair pairings and pay for
them with fractional paths and cycles removed before rounding. This preserves
the locality, degree bounds, and backbone visits required by Svensson's
algorithm and by recursive lifting (Theorem~\ref{thm:cover}).
The stronger cover reduces the cost of connecting quotient components;
passage-dependent accounting reduces the cost of completing their contracted
regions. The tour analysis combines these two improvements.

\subsection{From augmentation to \texorpdfstring{$(s,t)$}{(s,t)}-paths}
The augmentation guarantee keeps information that a complete-tour factor
does not express: the cost of completion decreases with the mass already
covered by the partial solution. Our transfer uses this information to
compare two initial walks before augmentation.

The standard normalization of the path LP supplies two inexpensive
$(s,t)$-paths in the support of its fractional solution
\cite[Lemmas~9.9, 9.10, and~9.16]{TV25}. One initial walk preserves both
path sections within every strongly connected component. The other visits
a maximum-coverage path from the fixed nice-path family in each component
and connects it to the entry and exit. The component maximum coverages bound
the added connector costs in the first construction; in the second, they
also certify mass that no longer needs completion. The first completion
bound increases with their sum, whereas the second decreases with it.
Keeping this shared quantity in both bounds and selecting the better
construction avoids the loss incurred by bounding each one separately.

The algorithm completes only the selected walk, with one augmentation call
per component. To apply Eulerian augmentation, it closes the walk by one
auxiliary return from $t$ to $s$. The component augmentations use only
internal arcs, so removing this return from the resulting Euler tour gives
a spanning $(s,t)$-walk. Section~\ref{sec:path} gives the two constructions
and proves the strengthened reduction to vertebrate pairs, then applies
the same analysis to our recursive augmentation.

\FloatBarrier

\section{Preliminaries}
\label{sec:preliminaries}
Parallel arcs and multiset unions are allowed, and all costs count multiplicity.
We discard self-loops. We call an arc multiset \emph{Eulerian} when indegree
equals outdegree at every vertex; connectivity is stated separately.
``Connected'' refers to the underlying undirected graph. A balanced connected multigraph is strongly
connected on its vertices and has an Euler tour. A single vertex with no arcs is an allowed connected Eulerian multigraph.

For a vertex set $S$, let $\delta^+(S)$ and $\delta^-(S)$ be its outgoing
and incoming arcs, and put $\delta(S)=\delta^+(S)\cup\delta^-(S)$.
Write $\diver x(v)=x(\delta^+(v))-x(\delta^-(v))$. The tour relaxation is
\begin{equation}
 \LP(G,c)=\min\{c^Tx:\ \diver x=0,\ x(\delta^-(S))\ge1
 \ (\emptyset\ne S\subsetneq V),\ x\ge0\}.
 \label{eq:tour-lp}
\end{equation}
For asymmetric path TSP with terminals $s,t$, replace balance by
$\diver x=\mathbf1_s-\mathbf1_t$ and impose
$x(\delta^+(S))\ge1$ for every nonempty $S\subseteq V\setminus\{t\}$.
Denote its optimum by $\LP_{s,t}(G,c)$.

\subsection{Strongly laminar instances and path coverage}
A \emph{strongly laminar instance} consists of a strongly connected digraph
$G$, a laminar family $\mathcal L$ of nonempty proper vertex sets, a
circulation $x>0$ feasible for~\eqref{eq:tour-lp}, and weights $y_L>0$.
Every $G[L]$ is strongly connected and $x(\delta^-(L))=1$. The induced
arc cost is $c_y(e)=\sum_{L:e\in\delta(L)}y_L$, so
$c_y^Tx=\sum_L2y_L=\LP(G,c_y)$.

For singleton weights, write $y_v=y_{\{v\}}$ when $\{v\}\in\mathcal L$
and $y_v=0$ otherwise. A \emph{vertebrate pair} consists of a strongly
laminar instance and a connected Eulerian backbone $B$ meeting every
nonsingleton laminar set. Following Traub and Vygen~\cite[Definition~7.7]{TV25} and
Vygen~\cite{Vyg26}, a $(\kappa,\eta)$-algorithm adds an Eulerian multiset $H$
such that $B\cup H$ is a spanning tour and
\begin{equation}
 c(H)\le\kappa\LP+\eta\sum_{v\notin V(B)}2y_v.
 \label{eq:vertebrate-interface}
\end{equation}
Thus $\kappa$ prices the LP value and $\eta$ prices uncovered singleton mass.
For $W\in\mathcal L\cup\{V\}$, a \emph{partial backbone} is a connected Eulerian
multigraph $B$ with $V(B)\cap W\ne\emptyset$, without the requirement to
meet all nonsingleton sets or to be contained in $W$. This is the more general
input in~\cite[Lemma~7.13]{TV25}. When a recursive construction chooses a new partial backbone,
we call it fresh to distinguish it from the inherited one.

\begin{proposition}[Laminar reduction and nice paths]
\label{prop:laminar}
The following operations are polynomial time~\cite[Theorem~7.4 and
Lemmas~7.5--7.6]{TV25}.
\begin{enumerate}[label=(\roman*)]
\item A ratio-$\rho$ algorithm for strongly laminar instances gives the
same LP-relative ratio for arbitrary nonnegative-cost ATSP. The reduction
restricts to an optimal primal support and supplies potentials with
$c_y(uv)=c(uv)+\pi_u-\pi_v$. Eulerian outputs on this support have the
same original and induced costs.
\item For $W\in\mathcal L\cup\{V\}$ and $u,v\in W$, one can find a
simple path $P^W_{uv}$ lying in their smallest laminar container and
entering and leaving every laminar set at most once.
\end{enumerate}
\end{proposition}
Following the terminology of~\cite[Lemma~7.5]{TV25}, call these paths
\emph{nice}. Fix these paths once and for all, allowing $P^W_{uu}$ to consist only of
$u$. Our augmentation theorem applies to any such fixed family.
For the classical bounds in Section~\ref{sec:path}, we use the global
choice of~\cite[Section~7.3]{TV25}: the same path for each endpoint pair
is used in every containing region, independently of the dual weights.
In both settings the caller and all recursive calls retain the chosen paths.
All subsequent arcs stay on the support, and we write $c=c_y$ until
the path normalization in Section~\ref{sec:path}.
For a region $W$, its descendant mass and vertex endpoint weights are
\begin{equation}
 M_W=\sum_{J\in\mathcal L:J\subsetneq W}2y_J,
 \qquad
 h_W(u)=\sum_{J\in\mathcal L:J\subsetneq W,\ u\in J}y_J.
\label{eq:geometry}
\end{equation}
For a vertex set $S$, define
\[
 \cov_W(S)=\sum_{\substack{J\in\mathcal L,\ J\subsetneq W\\J\cap S\ne\emptyset}}2y_J.
\]
For a path or subgraph $H$, write $\cov_W(H)=\cov_W(V(H))$.
The maximum coverage of a fixed nice path is
\[
 d_W=\max_{u,v\in W}\cov_W(P^W_{uv}).
\]
This maximum is over a polynomial-size path family. In the notation of
Traub and Vygen~\cite[Section~7.3]{TV25}, $M_W$ is $\operatorname{value}(W)$,
$\cov_W(P^W_{uv})$ is $D_W(u,v)$, and $d_W$ is $D_W$.
For a partial backbone $B$, $\cov_W(B)$ is their $D_W(B)$
from~\cite[Exercise~7.5]{TV25}.

\paragraph{Augmentation inside a region.}
\label{par:augmentation-output}
An \emph{augmentation of $B$ inside $W$} is an Eulerian arc multiset
$E_W$ supported on $E[W]$ such that $B\cup E_W$ is connected and visits
every vertex of $W$. The union retains every inherited copy of $B$;
the added cost is $c(E_W)$, even when an added arc also occurs in $B$.
The backbone may extend outside $W$, but every added arc lies inside it.
All algorithmic statements use rational input data, and polynomial time
includes the encoding length of the given backbone and the explicit size
of the added multiset.

\begin{lemma}[Nice-path coverage identity {\cite[Lemma~7.6]{TV25}}]
\label{lem:coverage}
For every fixed path $P^W_{uv}$,
\[
 h_W(u)+c(P^W_{uv})+h_W(v)=\cov_W(P^W_{uv})\le d_W\le M_W.
\]
Also $d_W=0$ if and only if $M_W=0$. Any subgraph containing this path
has coverage at least $\cov_W(P^{W}_{u v})$.
\end{lemma}
\begin{proof}
A visited set containing neither endpoint is crossed twice; a set containing
one endpoint contributes once through its boundary and once through an
endpoint weight. If both endpoints are in the set, the minimal-container
property keeps the path inside it, and the two endpoint terms contribute
its mass. An unvisited set contributes zero. Summing proves the identity.
If $M_W>0$, a vertex $u$ in a positive descendant has
$\cov_W(P^{W}_{u u})=2h_W(u)>0$.
\end{proof}

\subsection{Locality, witnesses, and the split graph}
An arc is \emph{local} if it crosses no nonsingleton laminar set
\cite[Definition~7.14]{TV25}. Thus $c(uv)=y_u+y_v$ for a local arc, and
for any local Eulerian multiset $H$,
\[
 c(H)=\sum_v \deg^-_H(v)\,2y_v.
\]
For a simple local circuit, every visited vertex has indegree one. This is
the vertex-cost identity used by Svensson's initialization and slack analysis.
A subtour cover for a local Eulerian $H$ outside a backbone $B$ is an Eulerian
multiset $F$ crossing every component of $H$, with every component of $F$
avoiding $B$ local~\cite[Definition~7.14]{TV25}. It need not be connected,
and its components meeting $B$ need not be local.

The witness flow and split graph enforce locality of components avoiding $B$
when rounding
\cite[Section~10.1]{STV20}. We recall the form of
Traub and Vygen~\cite[Definition~8.3 and Lemmas~8.4--8.5]{TV25}.
Order the nonsingleton sets in $\mathcal L\cup\{V\}$ by nonincreasing
cardinality, breaking ties consistently, as $L_1=V,L_2,\ldots,L_k$.
Set $r(v)=\max\{i:v\in L_i\}$. An arc $uv$ is forward, backward, or
neutral according as $r(u)<r(v)$, $r(u)>r(v)$, or $r(u)=r(v)$.
Neutral arcs are exactly the local arcs.
For a designated vertex set $B_0$, a witness is a vector $f$ satisfying
\begin{equation}
\begin{gathered}
 0\le f\le x,\qquad f_e=x_e\text{ on forward arcs},\qquad
 f_e=0\text{ on backward arcs},\\
 \diver f(v)\ge0\quad(v\notin B_0).
\end{gathered}
\label{eq:witness}
\end{equation}
The standard witness lemma is stated for a backbone meeting the nonsingleton
sets~\cite[Lemma~8.7]{TV25}. The following form records exactly the property
used by its proof.

\begin{lemma}[A hitting set suffices for a witness]
\label{lem:witness}
If $B_0$ meets every $L_i$, a witness exists and can be found in polynomial
time. The set $B_0$ need not induce a connected subgraph.
\end{lemma}
\begin{proof}
Let $C_i=\{v:r(v)=i\}$ and $S_i=\bigcup_{j\ge i}L_j$; then
$C_i\subseteq L_i\subseteq S_i$. Only the neutral variables in each
$C_i$ remain to be chosen. A capacitated flow feasibility test reduces to
\[
 x(V\setminus S_i\to U)\le x(U\to S_i\setminus U)
 \qquad(U\subseteq C_i\setminus B_0).
\]
Indeed, these are exactly the cut conditions for routing the prescribed
incoming forward flow through neutral arcs or outgoing forward arcs,
with unconstrained absorption at $B_0$.
For $i=1$ the left side is zero. For $i\ge2$, $L_i\setminus U$ is
nonempty because $B_0$ meets $L_i$. Subtracting
$x(\delta^-(L_i))=1$ from $x(\delta^-(L_i\setminus U))\ge1$ gives
$x(U\to L_i\setminus U)\ge x(V\setminus L_i\to U)$.
The displayed inequality follows by containment. Max-flow feasibility in
each rank class supplies the witness.
\end{proof}

Replace each vertex by lower and upper copies $v^0,v^1$. Forward arcs
occur only in the lower layer, backward arcs only in the upper layer,
and neutral arcs in both. Put $f_e$ on the lower copy and $x_e-f_e$ on
the upper copy. A downward switch $v^1v^0$ carries
$[\diver f(v)]_+$; an upward switch $v^0v^1$ carries
$[-\diver f(v)]_+$ and is permitted only at $v\in B_0$.
This is a circulation. Conversely its projection supplies a circulation
and a witness. Switches have cost zero. Every nonlocal split circuit uses
both layers and hence an upward switch at $B_0$. If the lower support is
acyclic, every circuit avoiding $B_0$ consists of neutral upper arcs.

\subsection{Algorithmic primitives}
We use integral bounded minimum-cost circulation, including ordinary
vertex gates, and exact rational linear programming from separation.
Finite integral lower and upper arc bounds ensure an integral optimum;
cost coefficients may be signed. Cut constraints below are separated by
directed minimum cuts. Integral circulation follows from~\cite[Theorem~3.11 and Corollary~3.12]{TV25};
optimization from separation follows from~\cite[Theorem~2.10]{TV25}
and~\cite{GLS88}. We use only the polynomial-time rational version.

The exchange arguments in Section~\ref{sec:cover} use the following
consequence of optimization from separation.

\begin{lemma}[Relative-interior point]
\label{lem:relative}
For a nonempty rational polytope with polynomially bounded rational data and a
polynomial-time separation oracle, a rational relative-interior point can be
computed in polynomial time. Every valid linear inequality tight there is an
equality on the entire polytope.
\end{lemma}
\begin{proof}
Starting with a feasible point, find the affine hull by optimizing basis
functionals orthogonal to the affine span of the points found so far. A
nonconstant functional supplies another feasible point outside that span.
After at most the dimension many extensions, the collected points span the
affine hull. Their average lies in no proper face: a valid inequality tight at
the average is tight at every collected point, hence on their affine span.
Optimization from separation~\cite{GLS88} and rational linear algebra give the
claimed complexity.
\end{proof}

\section{Boundary closure for recursive augmentation}
\label{sec:geometry}

Consider a recursive call on a proper laminar region $W$ with a partial
backbone $B$. Contracting the maximal descendants missed by $B$ exposes
an instance on which to run Svensson's algorithm, as in the recursive
reduction of Traub and Vygen~\cite[Lemma~7.13]{TV25}. There are two issues
before its subtour-cover subroutine can be used. The restriction of the
fractional circulation to $W$ need not be balanced, and a return through
an identified exterior vertex need not preserve attachment when expanded
into its prescribed path. We handle the cost and attachment issues here;
Section~\ref{sec:cover} proves the required rounding theorem for the
resulting auxiliary circulation.

Both transportation problems below are minimizations over fixed nice
paths. The reverse problem chooses paths from exits back to entries and
closes the restricted circulation. The forward problem gives endpoint
prices for passages through a child and a second way to choose a partial
backbone inside it. These uses are separate from the maximum transportation
problem in Section~\ref{sec:cover}, which will price uncertain repair
pairings after rounding.

\subsection{Boundary transportation and partial backbone choices}
For proper $W\in\mathcal L$, let
\[
 p^{\mathrm{in}}_u=x(\delta^-(W)\cap\delta^-(u)),\qquad
 p^{\mathrm{out}}_v=x(\delta^+(W)\cap\delta^+(v)).
\]
Both marginals have mass one. Allowed entry and exit ports are endpoints
of support arcs crossing $W$; later zero marginals do not authorize new
arcs. Let $\mathcal T(p^{\mathrm{in}},p^{\mathrm{out}})=\{Q\ge0:Q\mathbf1=p^{\mathrm{in}},Q^T\mathbf1=p^{\mathrm{out}}\}$.
Define
\begin{align}
 c_W^{\rightarrow}&=\min_{Q\in\mathcal T(p^{\mathrm{in}},p^{\mathrm{out}})}\sum_{u,v}Q_{uv}c(P^W_{uv}),&
 c_W^{\leftarrow}&=\min_{Q\in\mathcal T(p^{\mathrm{out}},p^{\mathrm{in}})}\sum_{v,u}Q_{vu}c(P^W_{vu}),\label{eq:transport}\\
 b_W&=\sum_up^{\mathrm{in}}_uh_W(u)+\sum_vp^{\mathrm{out}}_vh_W(v),&
 e_W&=c_W^{\leftarrow}-b_W.\label{eq:closure-value}
\end{align}
At the root set $e_V=0$. By Lemma~\ref{lem:coverage},
\begin{equation}
 c_W^{\rightarrow}+b_W\le d_W,\qquad c_W^{\leftarrow}+b_W\le d_W.
\label{eq:transport-bounds}
\end{equation}
The internal fractional cost is $M_W-b_W$, so adding reverse paths of
total cost $c_W^{\leftarrow}$ closes it at cost $M_W+e_W$.

\begin{lemma}[Bounded forward prices]
\label{lem:prices}
The forward transportation problem has an optimal dual $\varphi_u^{\mathrm{in}},\varphi_v^{\mathrm{out}}$ with
\[
 \varphi_u^{\mathrm{in}}+\varphi_v^{\mathrm{out}}\le c(P^W_{uv}),\qquad
 \sum_up^{\mathrm{in}}_u\varphi_u^{\mathrm{in}}+\sum_vp^{\mathrm{out}}_v\varphi_v^{\mathrm{out}}=c_W^{\rightarrow},\qquad
 -d_W/2\le \varphi_u^{\mathrm{in}},\varphi_v^{\mathrm{out}}\le d_W/2.
\]
The bounds hold at every allowed port, including zero-marginal ports.
\end{lemma}
\begin{proof}
Start with any optimal dual. Add a common constant to all entry prices
and subtract it from all exit prices until their maxima are equal.
Both marginals have mass one, so the objective is unchanged; each pair's
sum is unchanged as well. The constraint for a maximizing entry and exit
shows that both maxima are at most $d_W/2$. Raise every coordinate below
$-d_W/2$ to that value. Each affected pair then has sum at most zero,
so feasibility is preserved. The objective cannot decrease, and weak
duality preserves optimality.
\end{proof}

\begin{lemma}[Two partial backbones inside a region]
\label{lem:fresh}
Let $r>1$ be a coverage price, so a backbone's score is its cost minus
$r$ times its coverage. One can compute a connected Eulerian $B_W^{\max}\subseteq G[W]$
with cost at most $2d_W$, coverage at least $d_W$, and
\[
 c(B_W^{\max})-r \cov_W(B_W^{\max})\le(2-r)d_W.
\]
For proper $W$ one can also compute $B_W^{\rm tr}$ with
\[
 c(B_W^{\rm tr})-r \cov_W(B_W^{\rm tr})
 \le c_W^{\rightarrow}-(r-1)c_W^{\leftarrow}-rb_W.
\]
\end{lemma}
\begin{proof}
For the first choice concatenate $P^W_{uv}$ attaining $d_W$ with
$P^W_{vu}$. For the second, take an optimal forward coupling. The
partial backbone $P^W_{uv}+P^W_{vu}$ has score at most
\[
 c(P^W_{uv})-(r-1)c(P^W_{vu})-r(h_W(u)+h_W(v)),
\]
using the reverse path's coverage. The transpose coupling is feasible
for $c_W^{\leftarrow}$, so the average score has the asserted upper bound. Select
the best supported pair by its actual score.
\end{proof}

\subsection{Prescribed returns and attachment to the partial backbone}
Let $B$ be a partial backbone, that is, a connected Eulerian multigraph
meeting $W$, and let $\mathcal C$ be the inclusion-maximal positive proper
descendants of $W$ missed by $B$. This is the usual contraction step in
the recursive reduction~\cite[Lemma~7.13]{TV25}. The sets in $\mathcal C$
are disjoint, and every missed proper descendant lies in one of them. Then
\begin{equation}
 M_W=\cov_W(B)+\sum_{L\in\mathcal C}(2y_L+M_L).
\label{eq:mass-partition}
\end{equation}
Contract each $L\in\mathcal C$, keeping the original entry and exit
labels of every boundary arc. The quotient retains the induced laminar
structure. The image of $B$ meets every remaining nonsingleton descendant
of $W$. However, its portions inside $W$ need not be connected without
the excursions outside $W$.

An ordinary contraction of the exterior records only the total boundary
flow. We must also keep each exit--entry pairing and its prescribed return
path. We therefore use a separate exterior vertex for each pair with
positive coupling mass. This is what ``private'' means below.
The split graph will force a real visit to $B$ between consecutive
exterior visits, so the prescribed returns remain attached after lifting.

\begin{lemma}[Private exterior]
\label{lem:private}
For proper $W$ with $M_W>0$, choose an optimal reverse coupling $Q$.
For each $Q_{vu}>0$, introduce a private vertex $r_{vu}$ and arcs
$v r_{vu},r_{vu}u$, both of value $Q_{vu}$ and of combined cost
$c(P^W_{vu})$. Together with the original flow inside $W$, these arcs
form a circulation $x^W$. The same notation denotes its child contraction.
Let $V_{\mathrm{private}}$ be the private vertices and put
$B_0=(V(B)\cap W)\cup V_{\mathrm{private}}$ in the quotient. Then:
\begin{enumerate}[label=(\roman*)]
\item For every real set $S\subseteq W$ before contraction,
$x^W(\delta^-(S))=x(\delta^-(S))$. This also holds for quotient sets
and their preimages. Every child retains its unit boundary marginals.
\item Before child contraction the fractional cost is $M_W+e_W$.
Every integral private visit can be replaced at exactly its assigned cost
by the prescribed path $P^W_{vu}$.
\item The quotient has a witness for $B_0$ with each private incoming arc
upper and private outgoing arc lower. Private upward switches are forbidden.
Every split-walk segment between consecutive private visits therefore
contains a real vertex of $V(B)\cap W$.
\end{enumerate}
\end{lemma}
\begin{proof}
The coupling preserves every real vertex's incoming and outgoing boundary
amounts, proving balance and (i). Proper-descendant cut mass is $M_W$;
its contribution on arcs crossing $W$ is $b_W$. The internal flow
therefore costs $M_W-b_W$, and the private arcs add $c_W^{\leftarrow}$.
A private vertex has one entering and one leaving arc, so their integral
multiplicities agree, proving (ii).

For (iii), first contract $V\setminus W$ to a single ordinary exterior
vertex, and contract the children. Keep $W$ as a positive structural
laminar set and retain all descendants not strictly inside a child.
This is an ordinary strongly laminar quotient, with subtour-feasible
circulation. The projected partial backbone $B$ meets every retained nonsingleton set,
so Lemma~\ref{lem:witness} applies. There are at least two real quotient
vertices: otherwise $W$ would be a singleton or all of it a missed child,
contrary to $M_W>0$ and $B\cap W\ne\emptyset$.
Thus the structural set $W$ makes every exterior rank smaller than every
real rank. Split the exterior boundary according to $Q$. Its outgoing
arcs are forward and its incoming arcs backward; all witness constraints
at real vertices are unchanged. At $r_{vu}$, the downward switch has
value $Q_{vu}$ and the upward switch has value zero. These arc directions
force the same values in any later witness normalization. A segment
leaving a private vertex starts lower and arrives at the next private
vertex upper. It must use an upward switch at a real vertex of $B$.
\end{proof}

The private singletons may have fractional degree below one, so this
auxiliary graph is not an ordinary strongly laminar instance. Later
comparison prices can also give its private arcs signed costs. The cover
theorem accommodates these features by requiring cut inequalities only for
subsets of its real parts. For connectivity
tests the vertices of $B_0$ may be formally identified; this adds no usable
network arc. At the root there is no exterior: use the projected original
circulation and $B_0=V(B)$.

\section{A strengthened subtour cover}
\label{sec:cover}

The subtour cover used by Svensson's algorithm crosses every component of
the current local Eulerian solution. It may use nonlocal arcs, but every
component containing such an arc must meet the backbone. The guarantee used
by Vygen~\cite[Theorem~4]{Vyg26}, from Traub and Vygen~\cite[Theorem~8.24]{TV25}, is
\[
 c(F)\le 2\LP+\sum_{v\notin V(B)}2y_v,
 \qquad
 \deg_F^-(v)\le2\quad
 (y_v>0,\ \text{the component of $v$ avoids $B$}).
\]
The degree bound controls the local circuits charged to initialization
components; the total-cost bound pays for the terminal cover that attaches
them to the backbone.

We modify the same witness-flow construction. In the existing cover,
selected fractional paths crossing a current component $W_i$ are rerouted
through an auxiliary vertex, so that an integral visit there forces a
crossing of $W_i$. Restoring this vertex after rounding leaves a chosen
entry and exit that must be joined by a path inside $W_i$. These cover
auxiliaries are distinct from the private exterior vertices of
Section~\ref{sec:geometry}. The restoration path is local and simple, so
its cost is bounded by $\sum_{v\in W_i}2y_v$; summing over disjoint parts
produces the singleton surcharge~\cite[Section~8.5 and proof of
Theorem~8.24]{TV25}. We instead include a price for every possible
restoration in the rounding objective and pay for it with removed fractional
flow. Our target is $c(F)\le2c^Tx$ for the supplied circulation and costs,
before these internal modifications.

The recursive application also changes the required interface. The input
is the comparison circulation from Section~\ref{sec:geometry}, and its
cost need not be the optimum of a subtour LP. Contracted children are tight
vertices: the degree bound of four limits how often they must be expanded,
while the bound of two in components avoiding the designated set preserves
Svensson's local-circuit analysis. We retain the split-walk decomposition
and all visits to that set, so restoring private return paths preserves
attachment to the actual partial backbone. Finally, comparison prices may
make private arcs negative. Locality and the degree bounds are inherited
requirements: degree four appears in Svensson, Tarnawski, and
V\'egh~\cite[Lemma~10.3]{STV20}, and degree two away from the backbone in
Traub and Vygen~\cite[Theorem~8.24]{TV25}. We establish them together with
the improved cost and retained-visit guarantees in this more general
setting.

Let $x$ be a nonnegative rational circulation in a ranked directed graph, and let
$B_0$ be a designated vertex set. Let $W_1,\ldots,W_k$ partition $V\setminus B_0$,
where each $W_i$ induces a strongly connected graph and lies in one rank class.
Thus its internal arcs are neutral. A specified set $V_{\mathrm{tight}}\subseteq V\setminus B_0$
contains the vertices at which degrees must be bounded. As usual, degree means
indegree, which equals outdegree in an Eulerian multigraph. Let $V_{\mathrm{private}}\subseteq B_0$
be the set of private exterior vertices; every other vertex is real.

\begin{theorem}[Subtour cover]
\label{thm:cover}
Suppose that
\begin{enumerate}
\item $x(\delta^-(S))\geq 1$ for every nonempty $S\subseteq W_i$ and every $i$;
\item $x$ has a witness for $B_0$ in the sense of Lemma~\ref{lem:witness};
\item $x(\delta^-(v))=x(\delta^+(v))=1$ for every $v\in V_{\mathrm{tight}}$; and
\item $c$ is nonnegative on arcs with two real endpoints. Arcs incident to $V_{\mathrm{private}}$
may have signed costs.
\end{enumerate}
A polynomial-time algorithm returns an Eulerian multiset $F$ such that
\[
 c(F)\leq 2c^{\mathsf T}x,
 \qquad F\cap\delta(W_i)\neq\varnothing\quad(i=1,\ldots,k).
\]
Every component disjoint from $B_0$ is local. Each $v\in V_{\mathrm{tight}}$ has degree at most
four, and at most two if its component is disjoint from $B_0$. The algorithm
retains a decomposition into closed split walks. Every repair replaces a segment
inside one $W_i$ and preserves all visits to $B_0$ on that walk.
\end{theorem}

The proof keeps the split graph, acyclic witness, and doubling step of
Traub and Vygen~\cite[Chapter~8]{TV25}. A split vertex carrying at least
half a unit can force an integral visit directly. Otherwise an auxiliary
vertex collects half a unit of crossing walks, and rounding forces one
visit there. Restoring that visit requires an internal path whose endpoints
are chosen by the rounding. The difficulty is therefore to price every
possible entry--exit pairing, not just the original fractional pairing.

We first express this repair cost by a maximum transportation problem and
identify the fractional cost that can pay for it. We then normalize the
witness and select the crossing walks. When these walks can all pass through
a common vertex, their prefixes and suffixes may be paired at no additional
cost. The capped-flow argument supplies such a certificate while keeping
the actual rounded flow within the required degree bounds.

\subsection{Paying for restoration before rounding}

Write $f$ for a witness and $g=x-f$. In the split graph, $f$ uses the lower layer
and $g$ the upper layer. At $v\notin B_0$, the downward switch has value
$\diver f(v)\geq0$ and there is no upward switch. The upper throughput is
$g(\delta^-(v))$; the lower throughput is $f(\delta^+(v))$, including any flow
supplied by the downward switch. At a vertex of $V_{\mathrm{tight}}$, both are at most one.
We impose a throughput bound by splitting a node and adding an ordinary gate
arc. For a modified split circulation $z$, each arc has upper bound
$\lceil2z_e\rceil$, and each node gate with throughput $t$ has bounds
$\lfloor2t\rfloor$ and $\lceil2t\rceil$. Thus $2z$ is feasible, all bounds are
finite and integral, and minimum-cost circulation gives an integral solution of
no greater cost. Finite upper bounds allow signed objective coefficients.

Consider half a unit of selected split walks crossing a repair region contained
in one $W_i$. Erase cycles in their projections. When a repeated vertex occurs
first in the upper layer and then in the lower layer, replace the erased segment
by the downward switch at that vertex. This decreases original-arc flow,
preserves the entry and exit layers, and does not increase either copy's
throughput. Let $p^{\mathrm{in}}$ and $p^{\mathrm{out}}$ be the entry and exit marginals of the resulting
paths, and let $C_{\mathrm{path}}$ be their total internal cost. Redirect their boundary arcs
through an auxiliary vertex $v_{\mathrm{aux}}$. We use only two forms: all selected walks enter
in the upper layer, or all exit in the lower layer. In the first form, $v_{\mathrm{aux}}$ has no
lower original incoming arc and its upper throughput is $1/2$; in the second,
it has no upper original outgoing arc and its lower throughput is $1/2$.
Giving the designated gate bounds $1,1$ therefore forces exactly one original
incoming and outgoing arc at $v_{\mathrm{aux}}$. There is no upward switch at $v_{\mathrm{aux}}$.

For each allowed pair $(u,v)$, let $d_{uv}$ be the cost of a shortest simple
$u$--$v$ path inside the repair region. That region will be strongly connected,
and all its costs are nonnegative. Set
\begin{align}
 C_{\mathrm{repair}}&=\max\left\{\sum_{u,v}Q_{uv}d_{uv}:
              Q\geq0,\ Q\mathbf1=p^{\mathrm{in}},\ Q^{\mathsf T}\mathbf1=p^{\mathrm{out}}\right\}
 \label{eq:repair-transport}\\
  &=\min\left\{\sum_up^{\mathrm{in}}_u\varphi_u^{\mathrm{in}}+\sum_vp^{\mathrm{out}}_v\varphi_v^{\mathrm{out}}:
              \varphi_u^{\mathrm{in}}+\varphi_v^{\mathrm{out}}\geq d_{uv}\text{ for all }u,v\right\}.
 \nonumber
\end{align}
Add $\varphi_u^{\mathrm{in}}$ to the redirected incoming stub labelled $u$ and $\varphi_v^{\mathrm{out}}$ to the outgoing
stub labelled $v$. The fractional objective increases by $C_{\mathrm{repair}}$. Whatever pair the
integral circulation chooses, its two prices pay for the corresponding repair.
Individual prices may be negative; the circulation bounds remain finite.
This maximum transportation problem differs from both minimum transportation
problems in Section~\ref{sec:geometry}: those choose fractional boundary
routes, whereas this one covers the worst pairing that rounding may choose.

\begin{lemma}[Repair accounting]
\label{lem:repair}
Suppose that the construction cancels internal fractional cycles of total cost
$C_{\mathrm{cancel}}$ and removes path families of costs $C_{\mathrm{path},i}$ in disjoint parts, whose repair
transportation values are $C_{\mathrm{repair},i}$. After integral rounding and repair,
\[
 c(F)\leq 2c^{\mathsf T}x-2C_{\mathrm{cancel}}+2\sum_i(C_{\mathrm{repair},i}-C_{\mathrm{path},i}).
\]
\end{lemma}
\begin{proof}
The modified fractional objective is
$c^{\mathsf T}x-C_{\mathrm{cancel}}-\sum_iC_{\mathrm{path},i}+\sum_iC_{\mathrm{repair},i}$. Integral circulation rounding costs at
most twice this value. Each repaired auxiliary visit costs at most its two
already-paid endpoint prices by~\eqref{eq:repair-transport}.
\end{proof}

Modifications in distinct parts can share a boundary arc. This causes no
conflict. In either layer, if the arc has flow $z$ and selected tail and head
amounts $s,t\leq z$, choose
\[
 h\in[\max\{0,s+t-z\},\min\{s,t\}].
\]
Split the arc into four labelled classes of weights $h,s-h,t-h,z-s-t+h$, selected
at both ends, only the tail, only the head, or neither. Redirect the flagged
ends. Restoration retains the original arc once and pays each endpoint price
once. This permits all part modifications to be assembled before one rounding
operation.

We next normalize the witness. The primary boundary minimization is the one used
in the acyclic-witness construction of Traub and Vygen~\cite[Lemma~8.18]{TV25}.
We add a secondary cut optimization
to obtain the exchange properties needed for simultaneous modifications.
Lemma~\ref{lem:relative} lets us choose an optimum at which every tight
constraint is tight throughout the optimal face.

\begin{lemma}[Witness normalization]
\label{lem:normalization}
One can compute an acyclic witness $f$ minimizing
$\sum_i f(\delta(W_i))$. With its boundary values fixed, it maximizes, independently
in each part, the quantity
\[
 \mu_i=\min_{\varnothing\neq S\subseteq W_i}(x-f)(\delta^-(S)).
\]
For every inclusion-minimal minimizing shore $A_i$,
$g[A_i]$ is strongly connected and no positive-$f$ path inside $W_i$ leaves
$A_i$ and returns to it. Simultaneously contracting one such shore per part
preserves acyclicity of the support of $f$.
\end{lemma}
\begin{proof}
Minimize the boundary objective over the witness polytope and fix each resulting
boundary value. In each part maximize $\mu$ subject to
$(x-f)(\delta^-(S))\geq\mu$ for all its nonempty subsets. Contracting its
complement to a root and taking minimum root--vertex cuts separates these
constraints. The parts change disjoint sets of neutral arcs. Choose
relative-interior optima and combine them into $f_0$.

Finally minimize $\sum_e f_e$ subject to $f\leq f_0$, the primary optimum, and the
optimal cut constraints. A positive witness cycle uses only neutral arcs.
Deleting it preserves divergence and increases, rather than decreases, upper
cut capacities. If it decreased the primary boundary objective it would
contradict that objective's optimality. Otherwise deletion remains feasible in
the final program. Hence its optimum $f$ is acyclic. Boundary values only
decrease and their sum is fixed, so each remains fixed. A cut slack at $f_0$
remains slack; a cut tight at $f_0$ is an equality throughout the optimal face
and remains tight.

Fix one part and an inclusion-minimal minimum shore $A$. If $g[A]$ had several
strongly connected components, a source component would have incoming capacity
at most $g(\delta^-(A))=\mu$, contradicting minimality. Cut submodularity
also implies that any minimum shore intersecting $A$ contains $A$: otherwise
their intersection would be a smaller nonempty minimum shore.

Suppose that a positive-$f$ path inside the part leaves $A$ and returns. Decrease
$f$ along this excursion and increase it along a positive-$g$ path in $A$ with
the same endpoints. A sufficiently small change preserves box constraints,
divergence and boundary values. The internal path crosses no tight minimum
shore. Thus no tight upper cut decreases, while the incoming capacity of $A$
strictly increases. Small changes preserve all slack constraints, contradicting
the universal tightness of the cut of $A$.

For simultaneous contraction, a new cycle confined to one part would be such a
forbidden excursion. Any other new cycle consists of old positive witness
segments between selected shores. Its ranks are nondecreasing, so all segments
are neutral. Decrease $f$ along them and reconnect inside each selected shore
by a positive-$g$ path, from its outgoing port to its incoming port. This preserves
divergence and box feasibility for a small common amount. Internal changes do
not affect part boundaries, whereas some decreased segment crosses a boundary.
The primary objective strictly decreases, a contradiction.

An inclusion-minimal minimum shore is found by repeatedly testing whether a
minimum shore survives after deleting a vertex from the current shore. Each
test is another rooted minimum-cut problem; at most $|V|$ strict reductions
suffice.
\end{proof}

\subsection{Half a unit of walks through a common vertex}

The preceding accounting reduces the cost bound to finding crossing walks
whose removed cost, together with canceled cycles, pays for every repair.
We obtain this certificate from walks visiting a common vertex $w$:
prefixes and suffixes at $w$ can be paired in any way without changing their
combined cost. We first record a standard consequence of maximum flow.

\begin{lemma}[Walk packing]
\label{lem:packing}
Let $a$ be a nonnegative circulation with vertices $r,w$. If every $r$--$w$ cut
has capacity at least $1/2$, half a unit of closed $r$-walks visiting $w$ can be
packed within $a$.
\end{lemma}
\begin{proof}
Take an $r$--$w$ flow of value $1/2$. Subtracting it from $a$ leaves supply $1/2$
at $w$ and demand $1/2$ at $r$, so the residual contains a $w$--$r$ flow of that
value. Pair their path decompositions at $w$.
\end{proof}

Remove the root and retain the boundary labels. Artificial root arcs have cost
zero; original boundary arcs remain paid in the rounding objective. Re-pairing
prefixes and suffixes at $w$ realizes any entry--exit coupling. If cycle erasure
leaves paths of cost $C_{\mathrm{path}}$ and removes cycles of cost $C_{\mathrm{erase}}$, nonnegative internal
costs therefore give
\begin{equation}
 C_{\mathrm{repair}}\leq C_{\mathrm{path}}+C_{\mathrm{erase}}.
 \label{eq:hub-repair}
\end{equation}
Applying this packing directly may give too much flow through a tight
vertex. The next lemma separates the flow that will be rounded from an
internal circulation. The former obeys the degree caps; the latter supplies
closed detours to $w$ solely for the repair-cost certificate. Canceling that
circulation pays for the detours, without putting them into the rounded flow.

\begin{lemma}[Capped flow]
\label{lem:capped}
Let $g\geq0$ have nonpositive divergence at internal vertices of $W$, and assume
$g(\delta^-(S))\geq1/2$ for every nonempty $S\subseteq W$. Suppose
$0\leq g_0^{\mathrm{cap}}\leq g$ has the same boundary values and divergence as $g$, and
$g_0^{\mathrm{cap}}(\delta^-(v))\leq1/2$ at a specified set of capped vertices. For any $w\in W$,
one can compute $g^{\mathrm{cap}},g^{\mathrm{cyc}}\geq0$ such that $g^{\mathrm{cap}}+g^{\mathrm{cyc}}=g$, $g^{\mathrm{cyc}}$ is an internal circulation,
and $g^{\mathrm{cap}}$ respects the same caps, together with half a unit of boundary-originating
$g^{\mathrm{cap}}$-walks that all visit $w$ after inserting detours of combined arc usage at most
$g^{\mathrm{cyc}}$. Each detour returns to its insertion vertex.
\end{lemma}
\begin{proof}
Use a $g^{\mathrm{cap}}$ layer, a $g^{\mathrm{cyc}}=g-g^{\mathrm{cap}}$ layer, and a one-way switch from each $g^{\mathrm{cap}}$ copy to its
$g^{\mathrm{cyc}}$ copy, of capacity $1+\sum_e g_e$. A root supplies the $g^{\mathrm{cap}}$ layer through the
actual boundary entries; the target is the $g^{\mathrm{cyc}}$ copy of $w$. A minimum cut cannot
use a switch. Its two sink shores therefore satisfy $w\in S\subseteq T\subseteq W$,
where $S$ is in the $g^{\mathrm{cyc}}$ layer and $T$ in the $g^{\mathrm{cap}}$ layer. The minimum-cut value is
\[
 \mu=\min_{w\in S\subseteq T\subseteq W}
     \bigl\{g^{\mathrm{cyc}}(\delta^-(S))+g^{\mathrm{cap}}(\delta^-(T))\bigr\}.
\]
Optimize $g^{\mathrm{cap}}$ with its fixed boundary, divergence, box and throughput constraints
to maximize $\mu$. Layered minimum cuts give a separation oracle. Choose a
relative-interior optimum.

Suppose $\mu<1/2$. Minimum cuts are closed under intersection, so there is an
inclusion-minimal minimum pair $(S,T)$ contained in every minimum pair.
First, $g^{\mathrm{cyc}}[S]$ is strongly connected. Otherwise remove a source or sink strongly
connected component $K$ not containing $w$. For a source, circulation balance
shows that the incoming cut changes by $-g^{\mathrm{cyc}}(K,W\setminus S)$; for a sink it changes
by $-g^{\mathrm{cyc}}(W\setminus S,K)$. Keeping $T$ fixed would yield a smaller minimum pair.

Every throughput cap in $S$ is slack. Indeed, if a capped $v\in S$ has
$g^{\mathrm{cap}}$-indegree $1/2$, the boundary input $g^{\mathrm{cap}}(\delta^-(T))\leq\mu<1/2$ and the absence
of internal sources imply that a decomposition of $g^{\mathrm{cap}}[T]$ into simple
boundary-originating paths and cycles contains a positive cycle through $v$.
Move a small amount of this cycle from $g^{\mathrm{cap}}$ to $g^{\mathrm{cyc}}$. Caps only decrease. For every
tight pair $(S',T')$, we have $T\subseteq T'$, so its $g^{\mathrm{cap}}$ cut is unchanged and
its $g^{\mathrm{cyc}}$ cut cannot decrease. Small changes preserve slack constraints. The cap
at $v$ becomes slack at another optimum, contradicting relative interiority.

There is no positive-$g^{\mathrm{cap}}$ excursion from $S$ back to $S$ inside $T$. Otherwise
decrease $g^{\mathrm{cap}}$ along that excursion and increase it along a positive-$g^{\mathrm{cyc}}$ path in
$S$ with the same endpoints. Caps in $S$ are slack; elsewhere they can only
decrease. Each tight pair contains $S$ in its inner shore and $T$ in its outer
shore, so no tight cut capacity decreases. The capacity of $(S,T)$ strictly
increases, again contradicting universal tightness. Consequently no $g^{\mathrm{cap}}$ cycle
crosses $S$, and each boundary-originating path in $g^{\mathrm{cap}}[T]$ enters $S$ at most once.
Thus
\[
\begin{aligned}
 \mu&=g^{\mathrm{cyc}}(\delta^-(S))+g^{\mathrm{cap}}(\delta^-(T))\\
 &\geq g^{\mathrm{cyc}}(\delta^-(S))+g^{\mathrm{cap}}(\delta^-(S))\\
 &=g(\delta^-(S))\geq\tfrac12,
\end{aligned}
\]
a contradiction.

Take a connector flow of value $1/2$ in the layered network. Its $g^{\mathrm{cap}}$ pieces are
prefixes from boundary entries to switch vertices; let $s_v$ be the amount
switching at $v$. Complete $g^{\mathrm{cap}}$ to a circulation by root exits for boundary exits
and internal sinks. After removing the prefixes, the residual routes their
supplies $s_v$ to the root, completing them into boundary-to-boundary or
boundary-to-sink base walks within $g^{\mathrm{cap}}$. The connector's $g^{\mathrm{cyc}}$ pieces route
$s_v$ from the switch vertices to $w$. As $g^{\mathrm{cyc}}$ is a circulation, the residual
routes the same total amount from $w$ back to those vertices. Pair at $w$ by
insertion-vertex label. Each detour returns to the vertex where it was inserted,
and their combined usage is at most $g^{\mathrm{cyc}}$.
\end{proof}

\subsection{Constructing and rounding the cover}

Normalize $f$ by Lemma~\ref{lem:normalization}. Process each part $W$ using labelled
boundary stubs. First cancel upper cycles internal to $W$ on a probe copy of
$g$. If a lower throughput $f(\delta^+(v))$ or a probe upper throughput is at
least $1/2$, commit the probe deletions and force a visit to that split copy by
its gate. Both internal layers are now acyclic and there is no upward switch
in $W$. The forced visit must therefore cross $W$. No repair is needed.

Otherwise all lower and probe upper throughputs are below $1/2$. Define
$\mu=\min_{\varnothing\neq S\subseteq W}g(\delta^-(S))$ using the original,
uncanceled $g$. There are two cases.

\paragraph{A small upper cut.}
Suppose $\mu<1/2$, and let $A$ be an inclusion-minimal minimizing shore.
On a copy of $g[A]$, delete internal cycles, obtaining an acyclic flow with the
original boundary values and divergence. Keep the comparison circulation $x$
unchanged: this deletion reassigns flow to its lower complement. Reoptimize the
upper flow $\widetilde g$ inside $A$, with these boundary values and divergence fixed and
$0\leq \widetilde g\leq x$, to maximize
\[
 \min_{\varnothing\neq D\subseteq A}(x-\widetilde g)(\delta^-(D)).
\]
Choose a relative-interior optimum and make it acyclic by a decrease-only
minimization as in Lemma~\ref{lem:normalization}. Write $\widetilde f=x-\widetilde g$. Internal arcs
of $A$ are neutral, so this repartition remains a valid witness representation.

For a minimal minimum incoming shore $D$ of $\widetilde f$, the same exchange proof, with
$\widetilde g,\widetilde f$ replacing $f,g$, shows that $\widetilde f[D]$ is strongly connected and no $\widetilde g$ path
inside $A$ leaves $D$ and returns. The acyclic $\widetilde g$ flow has no internal sources
and total boundary input $\mu$. Its paths enter $D$ at most once, giving
\[
 \widetilde g(\delta^-(D))\leq\mu,
 \qquad \widetilde f(\delta^-(D))\geq1-\mu>\tfrac12.
\]
Every nonempty incoming cut of $\widetilde f$ in $A$ has the same lower bound.

Complete $\widetilde f[A]$ to a circulation with a root. Root entries represent actual
lower boundary entries and internal sources $\diver \widetilde f(v)\geq0$; exits represent
lower boundary exits. Added source arcs only increase rooted cut capacities.
Lemma~\ref{lem:packing} supplies half a unit of walks through any chosen vertex in
$A$. Extend walks starting at downward sources backward through $\widetilde g$ to actual
upper boundary entries of $A$. These extensions fit within $\widetilde g$, since their
selected demands are bounded by the downward-switch amounts, which are exactly
its internal sink amounts. Every extended walk exits $A$ in the lower layer.

Erase projected cycles, redirect the surviving paths through a lower-exiting
auxiliary, and then cancel all remaining lower cycles inside $A$.
Equation~\eqref{eq:hub-repair} and Lemma~\ref{lem:repair} show that the part
incurs no positive repair excess. The order matters: cycle cancellation occurs
\emph{after} rerouting, because the union of selected and unselected flow could
otherwise contain additional cycles.

These changes preserve global lower acyclicity. Boundary values on $A$ are
unchanged, and simultaneous contraction of the selected shores preserves
acyclicity by Lemma~\ref{lem:normalization}. Any new cycle would therefore lie
wholly inside one shore, where it has just been canceled. A cycle using its
auxiliary would project to a cycle through the contracted shore. Every rounded
visit to the lower-exiting auxiliary consequently lies on a split circuit
meeting $B_0$; its repair remains attached to that visit and crosses $W$.

\paragraph{No small upper cut.}
Now $g(\delta^-(S))\geq1/2$ for every nonempty $S\subseteq W$. The acyclic probe
is a feasible $g_0^{\mathrm{cap}}$ for Lemma~\ref{lem:capped}, with every vertex in $V_{\mathrm{tight}}\cap W$
capped at $1/2$. Apply that lemma. Extend base walks ending at internal upper
sinks through the acyclic lower flow to actual boundary exits of $W$. The
extensions fit because their demands are bounded by the downward switches.
All selected walks enter in the upper layer.

Let $C_{\mathrm{path}}$ be the cost of their base paths after cycle erasure, and $C_{\mathrm{erase}}$ the cost
erased. Inserting the $g^{\mathrm{cyc}}$ detours makes every walk visit the common vertex. Re-pairing there
therefore proves
\[
 C_{\mathrm{repair}}\leq C_{\mathrm{path}}+C_{\mathrm{erase}}+c(g^{\mathrm{cyc}}).
\]
Cancel $g^{\mathrm{cyc}}$, erase the selected base cycles, redirect only the remaining base
paths through an upper-entering auxiliary, and cancel remaining upper cycles
inside $W$. The detours certify the prices; they are not rerouted. Lemma~\ref{lem:repair}
again gives no positive repair excess. This auxiliary has no lower original
incoming arc and cannot create a lower cycle. Its unique upper incoming arc
comes from outside $W$, so its restored visit crosses $W$.

\paragraph{One rounding and the final guarantees.}
Assemble all modifications using the four-class split of boundary arcs. Apply
bounded minimum-cost integral circulation once, decompose its result into split
circuits with integer multiplicities, and restore each auxiliary's unique visit
by its priced simple path. The preceding cost inequalities sum to
$c(F)\leq2c^{\mathsf T}x$. Every part is crossed, and projection is Eulerian.

Before repair the lower support is acyclic. Thus a split circuit using it has
an upward switch and meets $B_0$. A circuit disjoint from $B_0$ stays in the upper
layer, where ranks never increase; it uses only neutral arcs. Any repair using
the lower layer was already on a circuit meeting $B_0$. Upper-to-upper repairs
remain upper. Hence every final component disjoint from $B_0$ is local.

In the case without an auxiliary, each designated split copy has throughput at
most one and rounds to at most two visits, giving degree four in total and two
in a component disjoint from $B_0$. In the small-cut case, lower throughputs
outside $A$ are below $1/2$, while upper throughputs are at most one. Inside
$A$, the acyclic upper flow has total input $\mu<1/2$, while lower throughput
is at most one. Thus the two rounded layers contribute at most three visits at
a designated vertex, and the unique simple repair adds at most one. That repair
belongs to a component meeting $B_0$, so disjoint components retain the upper
bound two. In the capped case, each layer contributes at most one and the repair
adds at most one, giving total degree at most three; a component disjoint from
$B_0$ has only the upper contribution and a possible upper repair, hence degree
at most two. The restored incoming boundary arc is part of this one added visit.

A repaired segment stays in its part and either preserves its layer or changes
once from upper to lower. In the latter case use a neutral upper path followed
by a downward switch at its exit. No repair introduces an upward switch outside
$B_0$, and none deletes a vertex of $B_0$. In particular, a real visit to $B_0$
between two private exterior visits remains present. This proves all assertions
of Theorem~\ref{thm:cover}.

All programs above have polynomially many variables and the stated minimum-cut
separation oracles. Cycle cancellation, flow decomposition, shortest paths and
integral circulation are polynomial-time procedures. Each part uses at most one
auxiliary, and boundary arcs split into only four classes per layer. Weighted
walks and integral multiplicities are stored in binary; the later normalization
of the primal circulation also bounds the explicit output size.

\section{Svensson's algorithm and recursive augmentation}
\label{sec:recursion}
We now use the strengthened subtour cover within Svensson's algorithm.
The algorithm joins connected components in a fixed quotient graph. After
its execution, we lift the chosen paths and recursively complete the
contracted regions. These are separate operations: the first changes an
Eulerian solution on one graph, while the second descends the laminar
hierarchy. The contraction and recursive architecture follow the
vertebrate-pair reduction of Traub and Vygen~\cite[Lemma~7.13]{TV25};
our extension uses the selected paths' coverage in the quotient algorithm's
charges and in the choice of the subsequent recursive completion.

\subsection{Svensson's algorithm with prescribed passages}
\label{subsec:merging}

Recall why the local-circuit analysis works in Vygen~\cite[Sections~2--4]{Vyg26}.
A local arc $uv$ has cost $y_u+y_v$, so a simple local circuit $C$ has
cost $\sum_{v\in V(C)}2y_v$. For a general local Eulerian multiset $K$,
the corresponding identity counts visits:
$c(K)=\sum_v\deg^-_K(v)\,2y_v$.
The vertex budgets pay for these costs. Slack measures the budget left
after paying for an initialization component, and a circuit either gives
a better initialization or can be charged to the components it meets.
The subtour cover supplies such circuits with bounded numbers of visits.
Its components meeting the backbone are paid only in the final iteration;
they need not be local.

After contraction, a vertex can represent a whole region $L$. A visit to
it must be expanded into a prescribed nice path $P$ between the
original entry and exit ports. By Lemma~\ref{lem:coverage}, its contribution
to a lifted local circuit is $2y_L+\cov_L(P)$, including the endpoint
weights on the incident arcs. The recursive reduction in Traub and Vygen~\cite[Algorithm~7.8]{TV25}
uses the fixed weight $2y_L+d_L$ at a contracted vertex. This pays for
every path in the fixed family, but discards the saving from a cheaper route. We retain
each route and extend the initialization and circuit-charging arguments
to its actual passage value. The fixed worst-case costs will be used only
to search for additional circuits.

Here the passage cost depends on the incoming and outgoing arcs together,
while the terminal cover uses a separate comparison objective. The following
theorem extends the initialization and charging argument to this setting.

Fix a contracted instance and a nonempty designated set $B_0$.
As in the private-return construction, $B_0$ represents attachments to one
actual partial backbone; it need not induce a connected subgraph.
Identify it as one component only for connectivity tests. This
identification adds no usable arc.  Write $U=V\setminus B_0$ and $n_0=|U|$.
For $v\in U$, let $y_v,d_v\geq0$ and $w_v=2y_v+d_v$.
Each passage through $v$ has a prescribed route $P$, with certified coverage
$\cov_v(P)\in[0,d_v]$.  In this abstract statement, these coverage values are supplied labels;
the recursive application uses the dual mass met by the nice path.
Routes and their entry and exit ports are fixed when a circuit is stored.
Assume that the cost of its lift is at most the sum of
$2y_v+\cov_v(P)$ over its passages. The recursive application will satisfy
this bound with equality for a local quotient circuit.
Give each local arc $uv$, with $u,v\in U$, the fixed search cost
$c_{\mathrm{raw}}(uv)=(w_u+w_v)/2$.

At each iteration, apply Theorem~\ref{thm:cover} with a fixed comparison
objective $c_{\mathrm{round}}$. The subtour-cover procedure returns an
Eulerian cover $F'$ crossing every
component of the current local solution, with
$c_{\mathrm{round}}(F')\leq2c_{\mathrm{round}}^{\mathsf T}x$.  Its components disjoint from $B_0$ are local,
and have indegree at most two at every vertex with $w_v>0$; the whole
cover has indegree at most four there.  We assume $c_{\mathrm{round}}$ is nonnegative on
local arcs.  We need no relation between $c_{\mathrm{round}}$ and $c_{\mathrm{raw}}$.

The parameter $\delta$ is a relative restart threshold: a restart must
gain more than $\delta$ times the budget outside the current initialization
component. It is chosen in terms of the requested approximation slack
$\varepsilon$ and controls the restart count and the resulting approximation loss.
The parameter $\gamma$ controls the initialization budgets, and $\theta$
sets the minimum coverage price of a retained passage.
Fix rational constants $\gamma>1$, $0<\delta<1$, and
$0<\theta\leq1$ such that
\begin{equation}
 (\gamma-1)\theta\geq1.
 \label{eq:merge-r}
\end{equation}
We use a floor $\theta d_v$ so that the search costs remain comparable
to the passage prices, even for routes of very small coverage. For a
stored passage, define
\[
 b_v(P)=2y_v+\max\{\theta d_v,\cov_v(P)\}.
\]
For a routed Eulerian multiset $K$, let $c_{\mathrm{pass}}(K)$ be the sum of these prices
over all its stored passages, and let $b_v(K)$ be their maximum at $v$.
For an empty singleton at $v$, set $c_{\mathrm{pass}}(K)=0$ and
$b_v(K)=2y_v+\theta d_v$.  Unions retain all passages with multiplicity;
therefore $c_{\mathrm{pass}}$ is additive, whereas $b_v$ is monotone under union.
We have
\begin{equation}
 c_{\mathrm{lift}}(K)\leq c_{\mathrm{pass}}(K),\qquad
 \theta c_{\mathrm{raw}}(K)\leq c_{\mathrm{pass}}(K)\leq c_{\mathrm{raw}}(K),\qquad
 \theta w_v\leq b_v(P)\leq w_v.
 \label{eq:merge-comparisons}
\end{equation}
For a simple circuit $C$, in particular,
$c_{\mathrm{pass}}(C)=\sum_{v\in V(C)}b_v(C)$.
For a multiset, the sum pays for every visit whereas the maximum supplies
one budget at each vertex; the two quantities are not equal in general.
This distinction will allow the final bound to be certified by an actual
component after lifting: the component containing a maximum-coverage
passage already covers at least as much mass as any other passage at that
child. We do not assume that the different lifted passages intersect.

\begin{theorem}[Svensson's algorithm with prescribed passages]
\label{thm:merge}
Under these assumptions, one can compute a local Eulerian multiset $H$
and a pruned subtour cover $F_{\mathrm{out}}$ such that
$B_0\cup H\cup F_{\mathrm{out}}$ is connected in the contracted graph.
The final cover satisfies the subtour-cover procedure's cost and degree bounds.
Its components meeting $B_0$ retain their original closed walks and
backbone visits; its other components retain their simple local circuits.
Let $b_v^{\mathrm{fin}}$ be the maximum price of a retained quotient
passage through $v$ in $H$ or the prescribed lift of $F_{\mathrm{out}}$,
using the singleton baseline if there is none.  Then
\begin{equation}
 c_{\mathrm{lift}}(H)\leq
 \alpha\sum_{v\in U}b_v^{\mathrm{fin}},\qquad
 \alpha:=\frac{\gamma}{1-\delta}
            +\frac{2\gamma}{\gamma-1}+\frac{\delta}{\theta}.
 \label{eq:merge-bound}
\end{equation}
For fixed parameters, the algorithm runs in polynomial time.
Passages inside separately paid exterior returns are not included in
$b_v^{\mathrm{fin}}$.
\end{theorem}

\paragraph{Initialization and the two restart tests.}
Assume first that $U\ne\emptyset$ and $w(U):=\sum_{v\in U}w_v>0$.
Give every vertex the same perturbation $\Delta=\delta w(U)/n_0$.
This supplies a positive minimum budget even at zero-weight vertices.  For a stored Eulerian multigraph $K$, define
\[
 \ell_K(S)=\sum_{v\in S\cap V(K)}
        \left(\frac{\gamma}{1-\delta}b_v(K)+\Delta\right),
 \qquad \ell(K)=\ell_K(V(K)),\qquad
 \sigma(K)=\ell(K)-c_{\mathrm{pass}}(K).
\]
Our $\delta$ plays the role of Vygen's restart tolerance $\varepsilon$,
and our budget coefficient $\gamma/(1-\delta)$ corresponds to his parameter
$\gamma$~\cite[Section~2.3]{Vyg26}.

As in Svensson's algorithm, the initialization is a local Eulerian
multiset $\widetilde H$. Write its connected components as $I_i$, with
vertex sets $W_i=V(I_i)$ partitioning $U$; empty singletons are allowed.
The additional stored information is the prescribed passage at every
visit. Unlike Vygen's fixed vertex budgets, $\ell_K$ can increase when
$K$ retains a passage of larger value. Monotonicity of these maxima is
what the restart proof below needs.
The components $I_i$ have nonnegative slack and are ordered so that
$\sigma(I_1)\geq\cdots\geq\sigma(I_k)$.
Set $W_0=B_0$.  The index of an object is the least index of an
initialization set it meets.  The initialization and its order remain
fixed throughout a stage.

Vygen's two tests examine a cover circuit itself and the improvement
obtained by adjoining it to its minimum-index initialization component
\cite[Lemmas~6--7 and Section~3.2]{Vyg26}. Here, for each surviving simple
local cover circuit $C$ of index $i>0$, test both $K=C$ and
$K=I_i\cup C$ for
\begin{equation}
 \sigma(K)>\sigma(I_i)
       +\delta\ell_K(V(K)\setminus W_i).
 \label{eq:merge-restart-test}
\end{equation}
If a test succeeds, let $J_{\mathrm{touch}}$ be the other initialization indices touched
by $K$, put $\ell_j=\ell_K(W_j\cap V(K))$, and retain the indices
$J_{\mathrm{keep}}=\{j\in J_{\mathrm{touch}}:\sigma(I_j)\geq \ell_j\}$.
Replace all touched objects by
$K^*=K\cup\bigcup_{j\in J_{\mathrm{keep}}}I_j$, and make newly uncovered vertices empty
singletons.  Untouched objects remain.  For $K=I_i\cup C$, the existing
copy of $I_i$ is used once.  The new object is connected because every
retained $I_j$ intersects $K$.

\begin{lemma}[Progress from a restart]
\label{lem:restart}
Starting with empty singletons, these restarts preserve nonnegative
slacks and occur polynomially many times, provided each tested circuit
crosses the current component containing its minimum-index object.
\end{lemma}
\begin{proof}
Write $\sigma_j=\sigma(I_j)$ and $\ell_{\mathrm{out}}=\sum_{j\in J_{\mathrm{touch}}}\ell_j$.
Monotonicity of the passage maxima gives
\[
 \ell(K^*)\geq\ell(K)+\sum_{j\in J_{\mathrm{keep}}}(\ell(I_j)-\ell_j).
\]
Subtracting the additive passage costs and using
\eqref{eq:merge-restart-test}, we obtain
\begin{equation}
 \sigma(K^*)>
 \sigma_i+\delta \ell_{\mathrm{out}}+
       \sum_{j\in J_{\mathrm{touch}}}\max\{\sigma_j-\ell_j,0\}
 \geq \sigma_i+\frac{\delta}{2}
                    \left(\ell_{\mathrm{out}}+\sum_{j\in J_{\mathrm{touch}}}\sigma_j\right).
 \label{eq:merge-slack-increase}
\end{equation}
The last inequality follows from
$(\delta/2)t+\max\{s-t,0\}\geq(\delta/2)s$ for $s,t\geq0$.
Define
\[
 \sigma_{\min}=\min_{v\in U}
       \left(\frac{\gamma}{1-\delta}(2y_v+\theta d_v)
                    +\Delta\right),\quad
 p=\left\lceil\frac{2}{\delta}\right\rceil,
 \quad \Phi=\sum_j\sigma_j^p.
\]
Every singleton has slack at least $\sigma_{\min}$, and
\eqref{eq:merge-slack-increase} preserves this lower bound.
Because the circuit crosses the current component containing $I_i$,
its outside budget satisfies $\ell_{\mathrm{out}}\geq \sigma_{\min}$.
Also $\sigma_i\geq\sigma_j$ for every touched $j$.
Convexity, $p\delta/2\geq1$, and $\sigma_i\geq \sigma_{\min}$ therefore give
\[
 \sigma(K^*)^p>
 \sigma_i^p+\sum_{j\in J_{\mathrm{touch}}}\sigma_j^p+\sigma_{\min}^p.
\]
The linear term in the convexity inequality contributes
at least $\sum_{j\in J_{\mathrm{touch}}}\sigma_j^p$ from the sum in
\eqref{eq:merge-slack-increase}, and at least $\sigma_{\min}^p$ from $\ell_{\mathrm{out}}$.
New singleton terms are nonnegative, so each restart increases $\Phi$
by more than $\sigma_{\min}^p$.
Since the initialization vertex sets are disjoint and $b_v(I_j)\leq w_v$,
\[
 \Phi\leq\left(\sum_j\sigma_j\right)^p
 \leq\left(\frac{\gamma w(U)}{1-\delta}+\delta w(U)\right)^p,
 \qquad \sigma_{\min}\geq\frac{\delta w(U)}{n_0}.
\]
The number of restarts is consequently bounded by
$[n_0(\gamma/(1-\delta)+\delta)/\delta]^p$.
Each restart retains a subset of the old circuits and adds one simple
circuit, so representations and rational arithmetic remain polynomial.
\end{proof}

\paragraph{One stage of the algorithm.}
Initialize $H=\widetilde H=\bigcup_i I_i$.
We retain the three operations of Svensson's algorithm: compute a subtour
cover, test for a better initialization, and add selected local circuits.
The procedure below either restarts or connects the quotient. No recursive
call is made during this execution.
\begin{enumerate}
\item Obtain a subtour cover $F'$.  Keep every component meeting $B_0$
intact, with its retained split walks.  Decompose the remaining components
into simple local circuits, reduce every positive circuit multiplicity
to one, and assign their nice routes.  Delete every such circuit
contained in one component of $B_0\cup H$; call the result $F$.
Apply both restart tests to every remaining local circuit.  If one
succeeds, restart with the new initialization and discard the other
objects from this stage.
\item Otherwise set $X=\emptyset$ and select a component $Z$ of
$B_0\cup H\cup F\cup X$ of largest index.  If its index is zero, return
$H\cup X$ and $F_{\mathrm{out}}=F$.
\item For $i=\operatorname{ind}(Z)>0$, search for a simple local circuit in
$G[U]$ crossing $Z$ with search cost at most $\frac{\sigma_i}{(\gamma-1)\theta}$.
If one exists, store it in $X$ with its prescribed routes, and repeat
the component selection in Step~2 without resetting $X$.
If none exists, add to $H$ all pieces of $F\cup X$ contained in $Z$,
discard the other temporary pieces, and return to Step~1.
\end{enumerate}

All subtour-cover crossing arcs survive pruning.  Thus every tested circuit
crosses a current component, as required by Lemma~\ref{lem:restart}.
Distinct components have distinct minimum initialization indices because
each $I_i$ is connected and retained in $H$.
A circuit crossing the selected largest-index component joins it to a
smaller-index component.  At most $n_0$ such insertions occur in a cover
call.  If the search fails, $Z$ contains at least two components of
$B_0\cup H$: otherwise the cover would not cross that component.
Retaining its pieces strictly reduces the number of current components.
There are therefore at most $n_0$ updates to $H$ in a stage.
The circuit search is polynomial: enumerate local arcs of $G[U]$ leaving
$Z$ and find a shortest return path using only local arcs of $G[U]$,
with objective $c_{\mathrm{raw}}$.
Nonnegative costs allow a simple shortest return path.

Every component retained before the terminal call avoids $B_0$ and contains only local
circuits.  Components of the cover meeting $B_0$ are retained only in
$F_{\mathrm{out}}$.  The returned union is connected, and pruning
preserves the comparison-cost and degree guarantees because it removes
only local circuits of nonnegative $c_{\mathrm{round}}$-cost.

\paragraph{Charging in the final stage.}
Consider the final stage, which has no restart.
For a surviving local cover circuit $C$ of index $i$,
failure of the first restart test gives
$\ell(C)-\delta\ell_C(V(C)\setminus W_i)-c_{\mathrm{pass}}(C)\leq\sigma_i$.
The budget expression before subtracting $c_{\mathrm{pass}}(C)$ is at least
$(1-\delta)\ell(C)\geq\gamma c_{\mathrm{pass}}(C)$.
Consequently,
\begin{equation}
 (\gamma-1)c_{\mathrm{pass}}(C)\leq\sigma_i,
 \qquad c_{\mathrm{raw}}(C)\leq \frac{\sigma_i}{(\gamma-1)\theta}.
 \label{eq:merge-cheap-cover}
\end{equation}
For the second test, the budget of $I_i\cup C$ on $W_i$ is at least
$\ell(I_i)$, and its outside budget equals that of $C$.
Failure therefore implies
\[
 (\gamma-1)\sum_{v\in V(C)\setminus W_i}b_v(C)
 \leq\sum_{v\in V(C)\cap W_i}b_v(C),
\]
whence
\begin{equation}
 c_{\mathrm{pass}}(C)\leq\frac{\gamma}{\gamma-1}\sum_{v\in V(C)\cap W_i}b_v(C).
 \label{eq:merge-anchor-charge}
\end{equation}

An initialization index receives cover charges in at most one
preterminal call.  Indeed, suppose a call retains a component $Z$ of
index $k$ containing a cover circuit of index $i\geq k$.
Then $W_i\subseteq Z$, and retention certifies that no local circuit in
$G[U]$ crossing $Z$ has search cost at most $\frac{\sigma_k}{(\gamma-1)\theta}$.
In a later call, a surviving circuit of index $i$ would meet $W_i$ but
leave its current $H$-component, which contains $Z$.
It would cross $Z$ with cost at most
$\frac{\sigma_i}{(\gamma-1)\theta}\leq \frac{\sigma_k}{(\gamma-1)\theta}$ by~\eqref{eq:merge-cheap-cover}, a contradiction.
Within its one charged call, the cover uses at most two passages at
each positive-price vertex.  Summing~\eqref{eq:merge-anchor-charge}
therefore bounds the preterminal cover cost by
\begin{equation}
 \frac{2\gamma}{\gamma-1}\sum_{v\in U}b_v^{\mathrm{fin}}.
 \label{eq:merge-cover-total}
\end{equation}

Charge a retained additional circuit from Step~3 to the index of its
selected component.  Its insertion joins that component to one of
smaller index.  If the circuit is retained, that union enters $H$, so
the charged index can never again be selected as a component minimum.
Within its call it already disappears upon insertion.
Thus each index pays for at most one retained additional circuit,
including those retained in the terminal call.
Together with the initialization, these circuits cost at most
\[
 \sum_i\bigl(c_{\mathrm{pass}}(I_i)+\frac{\sigma_i}{(\gamma-1)\theta}\bigr)
 \leq\sum_i\ell(I_i)
 \leq\frac{\gamma}{1-\delta}
             \sum_{v\in U}b_v^{\mathrm{fin}}+n_0\Delta.
\]
Here $(\gamma-1)\theta\geq1$, and every initialization passage is retained.
Finally,
$n_0\Delta=\delta w(U)
 \leq(\delta/\theta)\sum_v b_v^{\mathrm{fin}}$.
Combining this inequality with~\eqref{eq:merge-cover-total} and
\eqref{eq:merge-comparisons} proves~\eqref{eq:merge-bound}.

If $w(U)=0$, all raw and effective local costs vanish.
Set all budgets to zero, omit restarts, and use threshold zero in
Step~3; the same connectivity argument applies and the bound is zero.
If $U=\emptyset$, return $H=\emptyset$ and an unpruned subtour cover.
Keeping that cover also preserves its comparison bound when
$c_{\mathrm{round}}^{\mathsf T}x<0$.
This completes the proof of Theorem~\ref{thm:merge}.

As $\delta\downarrow0$, the coefficient approaches
$\gamma+2\gamma/(\gamma-1)$, whose minimum is $3+2\sqrt2$ at
$\gamma=1+\sqrt2$.  Taking $\theta=3/4$ satisfies
\eqref{eq:merge-r} for rational $\gamma$ sufficiently close to this value.

\begin{corollary}[Removing the singleton surcharge]
\label{cor:vertebrate}
For every fixed $\eta>3+2\sqrt2$, there is a $(2,\eta)$-algorithm
for vertebrate pairs in the sense of~\eqref{eq:vertebrate-interface}.
\end{corollary}
\begin{proof}
Choose fixed parameters for Svensson's algorithm for which the coefficient in
\eqref{eq:merge-bound} is at most $\eta$.
Apply Theorem~\ref{thm:merge} on the original graph with $B_0=V(B)$ and $d_v=0$ for
every $v\notin B_0$. The backbone meets every nonsingleton laminar set,
so all missed positive sets are singletons. Set
$V_{\mathrm{tight}}=\{v\notin B_0:y_v>0\}$; their singleton cuts are tight,
and every local arc $uv$ has cost $y_u+y_v$.
Thus $b_v^{\mathrm{fin}}=2y_v$, no nontrivial child path must be inserted,
and the original costs can serve as $c_{\mathrm{round}}$.
Theorem~\ref{thm:cover} bounds the terminal cover by $2\LP$, and
Theorem~\ref{thm:merge} bounds the additional local arcs by
$\alpha\sum_{v\notin V(B)}2y_v\le\eta\sum_{v\notin V(B)}2y_v$.
Their union with the backbone is a tour.
\end{proof}

\subsection{One recursive augmentation call}
The input now consists of a region $W$ and a partial backbone $B$: an
actual connected Eulerian multigraph meeting $W$, with no requirement to
meet all its positive descendants. Such recursive completion already
appears in the reduction to vertebrate pairs~\cite[Lemma~7.13]{TV25}.
Corollary~\ref{cor:vertebrate} supplies its standard $(2,\eta)$ input,
with the circuit-charging coefficient $\alpha$ bounded by $\eta$.
Our sharper recursion uses more of this algorithm's construction:
the charge in Theorem~\ref{thm:merge} depends on the retained passages'
coverage, which the $(2,\eta)$ guarantee alone does not specify.
We preserve those routes and use their coverage when comparing recursive
completions.

One call has the following order. Contract the inclusion-maximal positive
proper descendants of $W$ missed by $B$, retaining the entry and exit
ports of their boundary arcs. Every remaining nonsingleton laminar set
is met by $B$; otherwise it would lie inside one of the contracted sets.
For a proper $W$, close its fractional flow by the private return paths
of Lemma~\ref{lem:private}. Run the extension of Svensson's algorithm just
proved on this quotient, then expand its prescribed paths. Finally,
complete each contracted child by one recursive call. At this last step
we compare reuse of a lifted component with the two fresh partial
backbones of Lemma~\ref{lem:fresh}.

The quotient is connected when Svensson's algorithm stops, but its lift
can have several components: two visits to a contracted child may use
disjoint internal paths. Recursive completion spans that child and joins
all components incident with it. We prove this connectivity assertion
separately from the cost comparison below.

Let $\rho$ be the target tour approximation ratio. Choose fixed rational
parameters satisfying
\begin{gather}
 \gamma>1,\quad 0<\delta<1,\quad 0<\theta\le1,\quad
 \frac1{(\gamma-1)\theta}\le1,\qquad
 \alpha=\frac{\gamma}{1-\delta}+\frac{2\gamma}{\gamma-1}+\frac{\delta}{\theta},
 \label{eq:merging-parameters}\\
 \rho\ge10,\qquad \lambda=\frac1{\rho-3},\qquad
 \theta\le1-\frac2{\rho-2},\qquad \rho\ge\alpha+6+4\lambda.
 \label{eq:parameters}
\end{gather}
The coefficient $\lambda$ is the weight assigned to the forward
transportation prices in the comparison objective. In particular
$\rho\ge\alpha+2$ and $\lambda\le1$. The last condition closes the
recursive cost bound: the target ratio must satisfy
$\rho\ge\alpha+6+4/(\rho-3)$.

\begin{theorem}[Recursive augmentation]
\label{thm:augment}
Given a strongly laminar instance with its fixed nice paths, a region
$W\in\mathcal L\cup\{V\}$, and a partial backbone $B$ for $W$,
the recursive extension of Theorem~\ref{thm:merge} computes an augmentation
$E_W$ of $B$ inside $W$ satisfying
\begin{equation}
 c(E_W)\le\rho M_W+2e_W-(\rho-2)\cov_W(B).
 \label{eq:augmentation}
\end{equation}
For fixed rational parameters satisfying
\eqref{eq:merging-parameters}--\eqref{eq:parameters}, the algorithm is
polynomial time.
\end{theorem}

Since $e_W\le d_W$ for proper $W$ by~\eqref{eq:transport-bounds} and
$e_V=0$, the same output satisfies
\[
 c(E_W)\le\rho M_W+2d_W-(\rho-2)\cov_W(B).
\]
This is exactly the augmentation bound used in
Lemma~\ref{lem:path-completion} with $\kappa=2$; its parameter condition
$\rho\ge3\kappa+2$ follows from $\rho\ge10$.

We prove this by induction on the laminar hierarchy. The algorithm below
specifies which arcs are added; the child inequality proves their cost bound. If $M_W=0$, all
internal arc costs vanish. Join a vertex of $B\cap W$ to every vertex
of $W$ and back by support paths. Their union is a zero-cost Eulerian
augmentation. In this case $d_W=e_W=0$.
Suppose henceforth $M_W>0$. Form the children $\mathcal C$, the quotient,
and its private closure as in Section~\ref{sec:geometry}.

For each child $L$, define its fresh-backbone credit by
\begin{equation}
 \mathrm{credit}_L
 =\max\left\{\left(1-\frac2{\rho-2}\right)d_L,
       \frac{(\rho-3)(c_L^{\leftarrow}+b_L)-(c_L^{\rightarrow}+b_L)}{\rho-2}\right\}.
 \label{eq:child-prices}
\end{equation}
The first term comes from the maximum-coverage backbone; the second
uses the forward and reverse transportation costs. At least one of the
fresh backbones of Lemma~\ref{lem:fresh}, applied with $r=\rho-2$, has
cost minus $(\rho-2)$ times its coverage at most
$-(\rho-2)\mathrm{credit}_L$.
Thus this credit measures the cost-adjusted saving available from a
fresh backbone in the completion comparison below.
By~\eqref{eq:transport-bounds},
\begin{equation}
\begin{gathered}
 \left(1-\frac2{\rho-2}\right)d_L\le\mathrm{credit}_L\le\frac{\rho-3}{\rho-2}d_L,\\
 (1-\lambda)d_L\le(1+\lambda)\mathrm{credit}_L\le d_L.
\end{gathered}
 \label{eq:price-bounds}
\end{equation}
These formulas also apply when $d_L=0$, in which case
$c_L^{\rightarrow}=c_L^{\leftarrow}=b_L=\mathrm{credit}_L=0$ and we use zero endpoint prices.
The fixed search weight at a child is $2y_L+d_L$; at an uncontracted vertex
outside $B_0$ it is zero, since every missed positive singleton is a child.

Let $\varphi^{L,\mathrm{in}}_u,\varphi^{L,\mathrm{out}}_v$ be the bounded forward prices of Lemma~\ref{lem:prices}.
For a surviving quotient arc $e$, let $u_e$ be its original head label,
$v_e$ its original tail label, and let $c(e)$ retain its original cost
or its assigned share of a private return-path cost. Define the global
rounding objective by
\begin{equation}
\begin{split}
 c_{\mathrm{round}}(e)=c(e)
 &+\sum_{L\in\mathcal C:\,u_e\in L}
    \left(-h_L(u_e)+\frac{(1+\lambda)\mathrm{credit}_L}{2}-\lambda \varphi^{L,\mathrm{in}}_{u_e}\right)\\
 &+\sum_{L\in\mathcal C:\,v_e\in L}
    \left(-h_L(v_e)+\frac{(1+\lambda)\mathrm{credit}_L}{2}-\lambda \varphi^{L,\mathrm{out}}_{v_e}\right).
\end{split}
\label{eq:global-cost}
\end{equation}
Each sum has at most one term. On a real arc, an endpoint's remaining
contribution is at least $y_L+\bigl((1+\lambda)\mathrm{credit}_L-\lambda d_L\bigr)/2\ge0$.
Thus real arcs are nonnegative; private coefficients may be signed, as
permitted by Theorem~\ref{thm:cover}.

The exact fractional comparison value is
\begin{equation}
 c_{\mathrm{round}}^Tx^W=e_W+\cov_W(B)+
 \sum_{L\in\mathcal C}(2y_L+(1+\lambda)\mathrm{credit}_L-\lambda c_L^{\rightarrow}).
 \label{eq:comparison-value}
\end{equation}
Indeed, start with $M_W+e_W$. Contracting $L$ removes original internal
flow of cost $M_L-b_L$; the endpoint subtractions remove a further
$b_L$. Inflation contributes $(1+\lambda)\mathrm{credit}_L$, and the dual prices
subtract $\lambda c_L^{\rightarrow}$. Apply~\eqref{eq:mass-partition}.
The private paths are still paid in full: their internal arcs were not part
of the original internal flow removed during contraction.

We check the hypotheses of Theorems~\ref{thm:cover} and~\ref{thm:merge}. Current local
components are strongly connected and lie in one rank class. Every
positive-price quotient vertex outside $B_0$ is a tight child. A local
circuit's lift has cost equal to the sum of $2y_L+\cov_L(P)$ over its child
passages. The witness and real cut constraints follow from
Lemma~\ref{lem:private}, or from the original quotient at the root.
Apply Theorem~\ref{thm:merge}, with formal identification of $B_0$ only
for connectivity, to obtain local walks $H$ and a terminal cover $F_{\rm out}$.
Keep every route with its original entry and exit labels.

\paragraph{Lifting and selecting the recursive calls.}
A \emph{charged passage} is a visit to a child in a retained quotient
walk. Visits internal to a separately paid private return are excluded:
their path costs were already included in the comparison objective.
Expand the stored child passages and private returns, obtaining an Eulerian
multiset $K\subseteq E[W]$. Initially set $E_W=K$. For each child $L$,
fix a charged passage of maximum certified coverage. Process the children in a
fixed order. Just before processing $L$, let $B_L$ be the current actual
component containing that passage. Compare
\begin{equation}
 -(\rho-2)\cov_L(B_L),\qquad (4-\rho)d_L,\qquad
 c_L^{\rightarrow}-(\rho-3)c_L^{\leftarrow}-(\rho-2)b_L.
 \label{eq:three-scores}
\end{equation}
For the first choice, recursively augment $(L,B_L)$ and add only the
augmentation. For the second or third, build the corresponding fresh
backbone from Lemma~\ref{lem:fresh}, recursively augment it, and add both
the fresh backbone and its augmentation. Execute only the selected call.

\subsection{Connectivity and the recursive cost bound}
The connectivity argument has two parts. Completing the children joins
components by the elementary observation below. We then verify that the
lifted walks satisfy its hypothesis and supply the coverage needed for
the recursive cost bound.

\begin{lemma}[Joining components through disjoint regions]
\label{lem:joining}
Let $H$ be an Eulerian multigraph whose vertex set contains $W$, including
unvisited vertices of $W$ as isolated vertices, and let $\mathcal C$ be a
family of disjoint nonempty subsets of $W$. Form the bipartite graph with
one vertex for each component of $H$ and each $L\in\mathcal C$, joining
a component to a region when they intersect. Suppose this graph is connected.
Process the regions in any order, adding inside each $L$ an Eulerian
multiset that joins all current components meeting $L$.
The resulting multigraph is connected and Eulerian and spans $W$.
\end{lemma}
\begin{proof}
Adding arcs preserves balance and can only join components. Each initial
component of $H$ therefore lies in one final component. After processing
$L$, all initial components incident with $L$ lie in the same component,
and later additions preserve this fact. Connectivity of the incidence
graph implies that all initial components lie in one final component.
They include every vertex of $W$.
\end{proof}

\begin{lemma}[Actual components certify the lifted passages]
\label{lem:lifting}
The expanded multiset $K$ is Eulerian, and every retained walk of
$F_{\rm out}$ that met $B_0$ lifts to a walk meeting the actual $B$.
Each child has a charged passage. For the component $B_L$ selected above,
every charged passage $P$ through $L$ satisfies
\begin{equation}
 \cov_L(P)\le\min\{\cov_L(B_L),d_L\}.
 \label{eq:passage-coverage}
\end{equation}
If the completion of each child joins all current components meeting that
child and visits all its vertices, the final union with $B$ is connected
and spans $W$.
\end{lemma}
\begin{proof}
Every replacement uses a path with the same labeled endpoints, and hence
preserves the closed walk containing it. A split walk meeting a real
backbone vertex keeps that vertex. A walk visiting a private vertex has,
between consecutive private visits, a real backbone upward switch by
Lemma~\ref{lem:private}. The cover's repairs preserve that visit. Child
expansion cannot delete it because children miss $B$, and replacing a
private visit preserves both real endpoints. This proves attachment.

Every child lies outside the formal backbone, so quotient connectivity
gives a charged passage through it. The maximum such passage is contained
in one actual connected Eulerian component. Lemma~\ref{lem:coverage}
then gives~\eqref{eq:passage-coverage}; earlier additions only enlarge
this component and its coverage.

For connectivity, form a bipartite incidence graph whose vertices are the
actual components of $B\cup K$ and the children. Join a component to
each child it meets. Include unvisited vertices as isolated actual
components; such vertices lie inside children. Every quotient intersection
at an uncontracted real vertex is already an actual intersection. An
intersection at a child is represented by the corresponding incidence
vertex, and an attachment to $B_0$ is an attachment to the actual $B$.
The incidence graph is therefore connected. Lemma~\ref{lem:joining},
with $H=B\cup K$ and the unvisited vertices of $W$ included, now proves
the final assertion.
\end{proof}

Each choice in~\eqref{eq:three-scores} has the required joining property.
For reuse, induction connects and spans $L$ together with $B_L$, so it
intersects every existing component meeting $L$. A fresh completion is
itself connected and spans $L$, and has the same effect. Thus the algorithm
has the asserted connectivity.

For a terminal-cover passage entering $L$ at $u$ and leaving at $v$,
restoring the endpoint weights and adding its internal path changes the
paid cost by
\begin{equation}
 \cov_L(P^{L}_{u v})-(1+\lambda)\mathrm{credit}_L+\lambda(\varphi^{L,\mathrm{in}}_u+\varphi^{L,\mathrm{out}}_v)
 \le(1+\lambda)\bigl(\cov_L(P^{L}_{u v})-\mathrm{credit}_L\bigr).
 \label{eq:lift-correction}
\end{equation}
The dual constraint is for this same nice path. No correction is
needed for $H$, whose circuit-charging bound already measures its lift. Private
returns are inserted at their assigned full price. Any additional
coverage they create only helps; their internal visits need no terminal
degree charge.

\paragraph{The child inequality.}
We now combine the three costs associated with a child: its local
circuits, the terminal cover's prescribed passages, and its recursive
completion. A larger retained passage can cost more to lift, but it also
makes the reuse choice less expensive by increasing certified coverage.
The bound is largest where reuse becomes preferable to a fresh partial
backbone. This is the tradeoff captured in the following calculation.

Induction and Lemma~\ref{lem:fresh} bound the cost of completing $L$ by
\begin{equation}
 \begin{aligned}
 \rho M_L+2e_L+
 \min\bigl\{&-(\rho-2)\cov_L(B_L),(4-\rho)d_L,\\
            &c_L^{\rightarrow}-(\rho-3)c_L^{\leftarrow}-(\rho-2)b_L\bigr\}.
 \end{aligned}
 \label{eq:child-completion}
\end{equation}
The local-circuit charge at this child is at most
$\alpha(2y_L+\max\{\theta d_L,\min\{\cov_L(B_L),d_L\}\})$.
The terminal cover costs at most twice~\eqref{eq:comparison-value}
and has at most four charged passages through $L$. Therefore its lifting
excess is at most
$4(1+\lambda)[\min\{\cov_L(B_L),d_L\}-\mathrm{credit}_L]_+$.

The fresh-backbone credit was chosen so that
\[
 \begin{aligned}
 &\min\bigl\{-(\rho-2)\cov_L(B_L),(4-\rho)d_L,\\
 &\hspace{3em}c_L^{\rightarrow}-(\rho-3)c_L^{\leftarrow}-(\rho-2)b_L\bigr\}\\
 &\qquad\le -(\rho-2)\max\{\cov_L(B_L),\mathrm{credit}_L\}.
 \end{aligned}
\]
Indeed, the two fresh choices bound the minimum by
$-(\rho-2)\mathrm{credit}_L$; the transportation choice has a further
nonpositive term $-2b_L$. Also
\[
 2e_L-2\lambda c_L^{\rightarrow}
 \le2(c_L^{\leftarrow}+b_L)-2\lambda(c_L^{\rightarrow}+b_L),
\]
since $e_L=c_L^{\leftarrow}-b_L$ and $\lambda\le1$.
After removing the common term $\rho M_L$ and the boundary weight,
the remaining child charge is therefore bounded by the left-hand side
of~\eqref{eq:scalar}, with $A=\cov_L(B_L)$.

\begin{lemma}[Child charge]
\label{lem:scalar}
Under~\eqref{eq:parameters}, every child $L$ and every $A\ge0$ satisfy
\begin{equation}
\begin{split}
 2(1+\lambda)\mathrm{credit}_L
 &+\alpha\max\{\theta d_L,\min\{A,d_L\}\}\\
 &+4(1+\lambda)[\min\{A,d_L\}-\mathrm{credit}_L]_+\\
 &+2(c_L^{\leftarrow}+b_L)-2\lambda(c_L^{\rightarrow}+b_L)
   -(\rho-2)\max\{A,\mathrm{credit}_L\}\le0.
\end{split}
\label{eq:scalar}
\end{equation}
\end{lemma}
\begin{proof}
If $d_L=0$, the left-hand side is $-(\rho-2)A\le0$.
Otherwise $\theta d_L\le\mathrm{credit}_L<d_L$.
As a function of $A$, the left-hand side is nondecreasing on
$[0,\mathrm{credit}_L]$. On $[\mathrm{credit}_L,d_L]$ it has slope
$\alpha+6+4\lambda-\rho\le0$, and above $d_L$ it has slope $2-\rho$.
It is therefore largest at $A=\mathrm{credit}_L$, where it equals
\[
\begin{split}
 &2\bigl[(c_L^{\leftarrow}+b_L)-\lambda(c_L^{\rightarrow}+b_L)
                    -(1+\lambda)\mathrm{credit}_L\bigr]\\
 &\hspace{2em}-\bigl[\rho-\alpha-6-4\lambda\bigr]\mathrm{credit}_L.
\end{split}
\]
The first bracket is nonpositive by~\eqref{eq:child-prices} and
$1+\lambda=(\rho-2)/(\rho-3)$. The second bracket is nonnegative by the
parameter condition. This proves the inequality.
\end{proof}

Boundary mass $2y_L$ has coefficient $2+\alpha\le\rho$.
Summing the child charges leaves the parent term $2e_W+2\cov_W(B)$ and at most
$\sum_{L\in\mathcal C}\rho(2y_L+M_L)$.
By~\eqref{eq:mass-partition}, this is exactly the right side of
\eqref{eq:augmentation}. The connectivity argument above and the
algorithmic bounds below complete the proof of Theorem~\ref{thm:augment}.

\subsection{Polynomial running time and explicit output size}
\label{subsec:augmentation-runtime}
All transportation problems, witness programs, cut separations, shortest
paths, and bounded circulation optimizations use rational data of polynomial
encoding length. The relative-interior constructions in Section~\ref{sec:cover}
use polynomially many such optimizations. The initialization potential bounds
the number of restarts by a polynomial for fixed $\delta,\gamma,\theta$;
between restarts every update to $H$ reduces the number of quotient components.

We also need a polynomial bound on explicit arc multiplicities. Decompose
the original primal circulation into at most $|E|$ simple cycles and
truncate every coefficient above one to one. A truncated cycle crosses
no positive tight laminar cut, since otherwise its original contribution
to that unit cut would exceed one. Thus truncation preserves every tight
cut, its port marginals, and the induced objective. It preserves all other
subtour inequalities as well: either no truncated cycle crosses the cut,
or one retained unit cycle already supplies its required incoming value.
Balance and positive support are unchanged, and now $x_e\le|E|$.
Perform this normalization before computing the boundary data.

The private couplings have unit mass. Auxiliary constructions introduce
only polynomially many labeled arcs. Each rounding upper bound is the
ceiling of twice a polynomially bounded flow amount, so its integral output
has polynomially many arc copies. Each stored nice path has at most
$|V|-1$ arcs.

The recursive regions are original laminar sets. Distinct branches are
disjoint, and a branch descends strictly, so each laminar set is processed
at most once. Only one of the three choices is recursively executed.
Inherited backbone arcs are kept by reference, not reinserted. The
nonrecursive work of each call adds polynomially many new arcs,
independently of inherited multiplicities. Summing over $O(|V|)$ calls
proves polynomial total output size and running time.
The polynomial degree depends on the fixed approximation parameters.

\section{Tour and path guarantees}
\label{sec:guarantees}
We first choose the parameters for Theorem~\ref{thm:augment}. We then
prove one completion lemma that converts a recursive augmentation bound
into tour and path guarantees. Applying it to the reduction of Traub and
Vygen proves the transfer from vertebrate pairs; applying it to
Theorem~\ref{thm:augment} proves the common guarantee in
Theorem~\ref{thm:main}.

\subsection{Choosing the approximation parameters}
\label{subsec:parameter-choice}
As $\delta\downarrow0$ and rational $\gamma\to1+\sqrt2$, with
$\theta=3/4$, the circuit-charging coefficient tends to
$\alpha_*=3+2\sqrt2$. Equality in the decisive condition
$\rho\ge\alpha+6+4/(\rho-3)$ is
\[
 \rho^2-(\alpha_*+9)\rho+3\alpha_*+14=0.
\]
Its larger root is
\begin{equation}
 R_*=6+\sqrt2+\sqrt{15+6\sqrt2}.
 \label{eq:limiting-root}
\end{equation}
For any fixed $\varepsilon>0$, choose a rational $\rho$ with
$R_*<\rho<R_*+\varepsilon$, and set $\theta=3/4$ and
$\lambda=1/(\rho-3)$. Then $\rho>\alpha_*+6+4\lambda$.
The remaining inequalities involving $\gamma,\theta,\rho$ hold strictly
at $\gamma=1+\sqrt2$.
Taking rational $\gamma$ sufficiently close to this value and rational
$\delta>0$ sufficiently small therefore satisfies
\eqref{eq:merging-parameters}--\eqref{eq:parameters}.
These parameters depend only on $\varepsilon$, so the augmentation
algorithm runs in polynomial time for every fixed $\varepsilon>0$.
The completion lemma below supplies the tour and path guarantees.

\subsection{A stronger path reduction from vertebrate pairs}
\label{sec:path}
\label{subsec:prior-transfer}
The path reduction of Traub and Vygen~\cite[Theorem~9.23]{TV25}
converts a $(\kappa,\eta)$-algorithm into factor
$\max\{3\kappa+\eta+2,4\kappa+7\}$. Its second branch can prevent
an improvement in the vertebrate-pair algorithm from reaching paths.
For $\kappa\ge1$, the following theorem lowers that branch.

\begin{theorem}[Transfer from vertebrate pairs]
\label{thm:path-transfer}
For fixed rational $\kappa\ge1$ and $\eta\ge0$, a polynomial-time
$(\kappa,\eta)$-algorithm for vertebrate pairs gives a polynomial-time
LP-relative approximation for asymmetric path TSP with factor
\begin{equation}
 \max\left\{3\kappa+\eta+2,
       \;4\kappa+5-\frac{4\kappa+4}{2\kappa+\eta+2}\right\}.
 \label{eq:path-vertebrate}
\end{equation}
\end{theorem}

The recursive reduction of Traub and Vygen~\cite[Lemma~7.13]{TV25}
extends a partial backbone $B$ through a region $W$ at added cost at most
\[
 \kappa(M_W+d_W)
 +(2\kappa+\eta+2)\bigl(M_W-\cov_W(B)\bigr).
\]
This bound retains information that a complete-tour factor does not express:
completion becomes cheaper as $B$ visits more dual mass. We use this saving
to compare two initial $s$--$t$ walks. One preserves the coverage of two
inexpensive paths; the other visits a maximum-coverage nice path in each
strongly connected component of the positive LP support. Their completion
bounds depend oppositely on the same sum of
component coverage parameters. Combining them before eliminating that sum
proves Theorem~\ref{thm:path-transfer}.

We state the calculation in terms of the resulting tour factor $\rho$.
This also permits its use with the refined reduction of Traub and Vygen
and with our stronger recursive augmentation.

\begin{lemma}[Two-walk completion bound]
\label{lem:path-completion}
Let $\kappa\ge1$ and $\rho\ge3\kappa+2$ be fixed rational constants.
Suppose that, for every strongly laminar instance, one can choose a nice-path
family in polynomial time such that the following holds for every region
$W\in\mathcal L\cup\{V\}$ and partial backbone $B$ for $W$.
A polynomial-time procedure returns an augmentation $E_W$ of $B$ inside $W$
satisfying
\begin{equation}
 c(E_W)\le\rho M_W+\kappa d_W-(\rho-\kappa)\cov_W(B).
 \label{eq:transfer-interface}
\end{equation}
Here $d_W$ is defined using the selected family, fixed before $W$ and $B$.
Then there are polynomial-time LP-relative approximation algorithms with
tour factor $\rho$ and $(s,t)$-path factor
\begin{equation}
 \max\left\{\rho,\;4\kappa+5-\frac{4\kappa+4}{\rho-\kappa}\right\}.
 \label{eq:general-transfer}
\end{equation}
These factors agree whenever
\begin{equation}
 (\rho-\kappa)(\rho-4\kappa-5)+4\kappa+4\ge0.
 \label{eq:transfer-threshold}
\end{equation}
For $\kappa=2$, this condition is $\rho\ge(15+\sqrt{73})/2$.
\end{lemma}

The path algorithm constructs two initial walks and chooses between their
certified completion bounds. It then makes one augmentation call per
strongly connected component. The implementation of the augmentation
procedure is used only through~\eqref{eq:transfer-interface}.

\begin{proof}
For tours, Proposition~\ref{prop:laminar} reduces the problem to induced costs.
At the root, $B_V^{\max}$ from Lemma~\ref{lem:fresh} is a partial backbone
of cost at most $2d_V$ and coverage at least $d_V$.
Applying~\eqref{eq:transfer-interface} gives total cost at most
\[
 \rho M_V+(2\kappa+2-\rho)d_V\le\rho\LP,
\]
since $\rho\ge3\kappa+2\ge2\kappa+2$. In the application to
Theorem~\ref{thm:augment}, its sharper bound with $e_V=0$ gives
\begin{equation}
 \rho M_V+(4-\rho)d_V\le\rho\LP.
 \label{eq:tour-bound}
\end{equation}
Vertex potentials cancel on the Eulerian output, and metric shortcutting
does not increase cost.

\medskip\noindent\textbf{Normalization and component completion.}
\label{subsec:path-normalization}
For distinct terminals $s,t$, recall the path relaxation $\LP_{s,t}$
from Section~\ref{sec:preliminaries}. We work on the positive support of an optimum
solution $x$.
The following facts are the path normalization and two-path construction
in the framework of K\"ohne, Traub, and Vygen~\cite{KTV20}, in the form of Traub and Vygen~\cite[Lemmas~9.9, 9.10, and~9.16]{TV25}.
In polynomial time one obtains a laminar family
$\mathcal L\subseteq 2^{V\setminus\{s,t\}}$, positive weights $y_L$,
and vertex potentials $\pi$, such that each $G[L]$ is strongly connected,
$x(\delta^-(L))=x(\delta^+(L))=1$, and
\begin{equation}
\begin{gathered}
 c_y(uv)=\sum_{L:\,uv\in\delta(L)}y_L
        =c(uv)+\pi_u-\pi_v,\\
 \LP_{s,t}=M-\pi_s+\pi_t,
 \qquad M=\sum_{L\in\mathcal L}2y_L.
\end{gathered}
\label{eq:path-potentials}
\end{equation}
The potentials are chosen to minimize $\pi_s-\pi_t$ among optimal dual solutions.
There are support $s$--$t$ paths $P_1,P_2$ whose combined number of
crossings of each positive laminar cut is at most two.
Consequently, the paths never both visit a set of $\mathcal L$, and their
total induced cost is the mass of the sets visited by their union.
If $M>0$, define $\tau\in[0,1]$ by
\begin{equation}
 c_y(P_1)+c_y(P_2)=\tau M.
 \label{eq:path-tau}
\end{equation}
Since the original costs are nonnegative,
$0\leq c(P_1)+c(P_2)=\tau M-2(\pi_s-\pi_t)$.  Therefore
\begin{equation}
 M\leq \frac{2\LP_{s,t}}{2-\tau}.
 \label{eq:path-mass-bound}
\end{equation}

Let $W_1,\ldots,W_k$ be the strongly connected components of the support,
in topological order.  All intercomponent arcs go from $W_i$ to $W_{i+1}$,
and their total $x$-value is one; also $s\in W_1$ and $t\in W_k$.
Every $P_j$ thus has a nonempty section in every component.
Every positive laminar set is contained in a component because its induced
graph is strongly connected.  It cannot equal that component: both paths
would then visit it.  Thus every such set is a \emph{proper} subset of one
$W_i$.

To invoke the augmentation procedure, add a vertex $r$ and arcs $tr,rs$,
each with unit flow.  This closes $x$ to a feasible subtour circulation.
Adjoin the sets $W_i$ to the laminar family and give each a positive own
weight, say one, solely for this invocation. These weights contribute only
to arcs crossing component boundaries.
Choose the family supplied by the hypothesis for this augmented instance
and retain its paths inside each $W_i$ in every invocation.
Define
\begin{equation}
 M_i:=M_{W_i},\qquad d_i:=d_{W_i},\qquad
 \sum_i M_i=M,\qquad D:=\sum_i d_i\leq M.
 \label{eq:path-components}
\end{equation}
For this fixed family, $M_i,d_i,\cov_{W_i}(Q)$ and internal arc costs
are the same under the original and augmented weights.
The initial walks and the augmentation bound use these same $d_i$.
For every support $s$--$t$ walk $Q$, the multigraph
$B=Q\cup\{tr,rs\}$ is connected and Eulerian. The assumed procedure
returns Eulerian arcs $E_i\subseteq E[W_i]$ such that $B\cup E_i$ is
connected and spans $W_i$, with
\begin{equation}
 c_y(E_i)\le\rho M_i+\kappa d_i-(\rho-\kappa)\cov_{W_i}(Q).
 \label{eq:path-completion}
\end{equation}
Every call uses the same inherited $B$.
All costs in the rest of this section are evaluated with the original
weights in~\eqref{eq:path-potentials}, not the temporary component weights.

The union of $B$ and all $E_i$ is connected, spanning, and Eulerian.
Since the $E_i$ are internal to the original components, the auxiliary
vertex has exactly one entering and one leaving arc.
Cutting an Euler tour there and deleting $tr,rs$ gives a spanning
$s$--$t$ walk $F$ with
\begin{equation}
 c(F)=c_y(Q)+\sum_i c_y(E_i)-\pi_s+\pi_t.
 \label{eq:path-final-cost}
\end{equation}
If $M=0$, all internal induced costs vanish.  An arbitrary support
$s$--$t$ path, completed by zero-cost closed walks within its components,
already has cost $\pi_t-\pi_s=\LP_{s,t}$.  We henceforth assume $M>0$.

\medskip\noindent\textbf{Two complementary initial walks.}
\label{subsec:path-walks}
The two constructions trade the cost of visiting each component against
the mass it leaves uncovered.

\noindent\textbf{Preserving both paths.}
Inside odd-indexed components, traverse the section of $P_1$ followed
by that of $P_2$; reverse this order in even-indexed components.
Join the sections by one nice path.  Between consecutive
components, use the intercomponent arc of the path traversed last.
The order alternation makes the concatenation a support $s$--$t$ walk,
which we denote by $Q_{\mathrm{keep}}$.
It retains all vertices of $P_1\cup P_2$ and has
\[
 c_y(Q_{\mathrm{keep}})\leq\tau M+D,
 \qquad \sum_i \cov_{W_i}(Q_{\mathrm{keep}})\geq\tau M.
\]
Applying~\eqref{eq:path-completion} and~\eqref{eq:path-final-cost},
the completion of $Q_{\mathrm{keep}}$ has cost at most
\begin{equation}
 U_{\mathrm{keep}}:=\LP_{s,t}
       +\bigl[\rho-1-(\rho-\kappa-1)\tau\bigr]M+(\kappa+1)D.
 \label{eq:path-F2}
\end{equation}

\noindent\textbf{Replacing the path sections.}
Let $s_i,t_i$ be the first and last vertices of $P_1$ in $W_i$.
Choose $u_i,v_i\in W_i$ with $\cov_{W_i}(P^{W_i}_{u_i v_i})=d_i$.
Inside $W_i$, traverse the three nice paths
\[
 P^{W_i}_{s_i u_i},\qquad
 P^{W_i}_{u_i v_i},\qquad
 P^{W_i}_{v_i t_i}.
\]
Use the $P_1$ arc $t_i s_{i+1}$ between successive components.
This defines another support $s$--$t$ walk, denoted by $Q_{\mathrm{max}}$.

Write $h_i=h_{W_i}$.  The nice-path identity gives the internal
cost bound
\[
 3d_i-h_i(s_i)-2h_i(u_i)-2h_i(v_i)-h_i(t_i).
\]
Every original positive laminar set is strictly inside one component,
so the intercomponent arc has cost
\[
 c_y(t_i s_{i+1})=h_i(t_i)+h_{i+1}(s_{i+1}).
\]
These endpoint terms telescope, and
$h_1(s)=h_k(t)=0$ because the laminar sets exclude the terminals.
Hence
\begin{equation}
 c_y(Q_{\mathrm{max}})\leq
 3D-2\sum_i\bigl(h_i(u_i)+h_i(v_i)\bigr)\leq3D.
 \label{eq:path-Q3}
\end{equation}
The path $P^{W_i}_{u_i v_i}$ certifies coverage at least $d_i$ in $W_i$.
Its completion therefore has cost at most
\begin{equation}
 U_{\mathrm{max}}:=\LP_{s,t}+(\rho-1)M-(\rho-2\kappa-3)D.
 \label{eq:path-F3}
\end{equation}

Figure~\ref{fig:path-walks} shows the two constructions inside consecutive
components.
\PathWalksFigure

\medskip\noindent\textbf{Combining the guarantees.}
\label{subsec:path-combination}
Both bounds are known before augmentation. We choose the initial walk
with the smaller certified bound and complete only that walk, obtaining
$F$ with $c(F)\le\min\{U_{\mathrm{keep}},U_{\mathrm{max}}\}$.
Since $\rho\ge3\kappa+2$ and $\kappa\ge1$, the weights
$(\rho-2\kappa-3)/(\rho-\kappa-2)$ and $(\kappa+1)/(\rho-\kappa-2)$ are nonnegative and sum
to one. Applying them to~\eqref{eq:path-F2} and~\eqref{eq:path-F3},
respectively, cancels $D$ and gives
\begin{equation}
 c(F)
 \leq \LP_{s,t}
       +\left(\rho-1-\frac{(\rho-2\kappa-3)(\rho-\kappa-1)}{\rho-\kappa-2}\tau\right)M.
 \label{eq:path-combined}
\end{equation}
Also $D\leq M$ in~\eqref{eq:path-F2} gives
$c(F)\leq \LP_{s,t}+(\rho+\kappa-(\rho-\kappa-1)\tau)M$.
Both coefficients of $M$ are nonnegative on $0\le\tau\le1$:
the first has minimum $2\kappa+1+(\kappa+1)/(\rho-\kappa-2)>0$, and
the second has minimum $2\kappa+1$. Thus
\eqref{eq:path-mass-bound} proves the ratio bound
\begin{equation}
 1+\frac{2}{2-\tau}
       \min\left\{\rho-1-\frac{(\rho-2\kappa-3)(\rho-\kappa-1)}{\rho-\kappa-2}\tau,
                   \ \rho+\kappa-(\rho-\kappa-1)\tau\right\}.
 \label{eq:path-envelope}
\end{equation}
The two ratio bounds in~\eqref{eq:path-envelope} meet at
$\tau=(\rho-\kappa-2)/(\rho-\kappa-1)$, and the first is smaller
before that point. Its derivative has sign opposite to
$(\rho-\kappa)(\rho-4\kappa-5)+4\kappa+4$. The second branch is nonincreasing
because $\rho\ge3\kappa+2$. Hence the maximum of their lower envelope
occurs at $\tau=0$ or at the meeting point. The two values are
$\rho$ and $4\kappa+5-(4\kappa+4)/(\rho-\kappa)$, proving
\eqref{eq:general-transfer}. Comparing them gives
\eqref{eq:transfer-threshold}.

The fixed nice-path families and their maximum-coverage choices are computable in
polynomial time. The selected walk invokes the augmentation algorithm
once per component, and the components are disjoint. Thus the algorithm
is polynomial time whenever the augmentation algorithm is.
\end{proof}

\paragraph{Obtaining an internal augmentation from the classical bounds.}
\label{par:classical-internalization}
The regional conclusions of Traub and Vygen~\cite[Lemma~7.13 and
Exercises~7.5--7.6]{TV25} give an Eulerian addition $F$ such that
$B\cup F$ is a tour on $W\cup V(B)$, together with a bound on $c_y(F)$.
We justify the internal-support condition needed in
Lemma~\ref{lem:path-completion} directly from these conclusions.
Both cost bounds depend only on $\kappa,\eta$, the refinement parameter,
and the proper-descendant quantities $M_W,d_W,\cov_W(B)$.

For proper $W$, let $R\ge0$ denote the applicable source bound and replace
only the weight of $W$ by
\[
 y'_W=y_W+R+1.
\]
Use the global nice paths fixed in~\cite[Section~7.3]{TV25} before this
change; their choice depends only on the graph and laminar family.
The same primal circulation remains
feasible and tight, so the modified instance is strongly laminar.
All proper-descendant quantities and all internal arc costs in $W$ remain
unchanged. Apply the cited regional guarantee to this modified instance.
Its output satisfies $c_{y'}(F)\le R$. Every arc crossing $W$ has cost
at least $y'_W>R$, so none can occur in $F$.

Set $E_W=F\cap E[W]$, retaining multiplicities. Since $F$ has no boundary
arc, $E_W$ is Eulerian. For a vertex of $W$ outside $B$, follow a path in
$B\cup F$ to its first vertex of $B$. Before that vertex the path uses
only $F$, hence remains inside $W$ and is retained in $E_W$.
Thus $B\cup E_W$ is connected and spans $W$, while
$c_y(E_W)=c_{y'}(E_W)\le c_{y'}(F)\le R$.
For $W=V$ the source output is already internal.
The weight change and restriction take polynomial time and preserve
polynomial encoding length for rational input data.

\begin{proof}[Proof of Theorem~\ref{thm:path-transfer}]
Traub and Vygen's recursive bound~\cite[Lemma~7.13]{TV25}, localized
as above, satisfies~\eqref{eq:transfer-interface} with
$\rho=3\kappa+\eta+2$. Substitute this value in
Lemma~\ref{lem:path-completion} and~\eqref{eq:transfer-threshold}.
The path factor agrees with the standard tour factor $3\kappa+\eta+2$
whenever $\eta^2+(\kappa-1)\eta-2(\kappa+1)^2\ge0$.
\end{proof}

\subsection{The refined reduction and the new augmentation}
For $\kappa=2$, the previous transfer agrees with the standard tour
factor only for $\eta\ge7$. Theorem~\ref{thm:path-transfer} lowers
this threshold to $(\sqrt{73}-1)/2$. In particular, it applies to
Vygen's $(2,4+2\sqrt2+\varepsilon)$-algorithm~\cite[Theorem~3]{Vyg26}.
The refined recursive reduction used for his tour bound gives a stronger
consequence. We express its guarantee directly in terms of the tour
factor $\rho$.

\begin{lemma}[Refined recursive reduction {\cite[Exercises~7.5--7.6]{TV25}}]
\label{lem:prior-refined-augmentation}
Given a polynomial-time $(\kappa,\eta)$-algorithm for vertebrate pairs,
where $\kappa\ge1$ and $\eta\ge0$ are fixed rational constants,
let $\rho>\kappa+2$ be a fixed rational
constant satisfying
\begin{equation}
 \rho\ge3\kappa+\eta+2-\frac{2}{\rho-\kappa}.
 \label{eq:prior-refined-rho}
\end{equation}
For every strongly laminar instance, one can choose a nice-path family in
polynomial time such that, for every region $W\in\mathcal L\cup\{V\}$
and partial backbone $B$ for $W$, a polynomial-time procedure returns an
augmentation of $B$ inside $W$ with cost bounded by
\eqref{eq:transfer-interface}. The family is fixed before $W$ and $B$;
$d_W$ is defined using this family.
\end{lemma}
\begin{proof}
Set $\delta=2/(\rho-\kappa)$. Since $\rho>\kappa+2$, we have
$0<\delta<1$. Condition~\eqref{eq:prior-refined-rho} gives
\[
 2\kappa+\eta+2-\delta\le\rho-\kappa,
 \qquad
 (2\kappa+\eta+1-\delta)\delta\le2-\delta,
\]
where the second inequality follows by multiplying the first by $\delta$
and using $(\rho-\kappa)\delta=2$.
Hence Traub and Vygen's refined regional guarantee
\cite[Exercises~7.5--7.6]{TV25} applies and bounds the added cost by
\[
 \kappa(M_W+d_W)
 +(2\kappa+\eta+2-\delta)(M_W-\cov_W(B)).
\]
Localize this output by the preceding weight modification. Since
$M_W-\cov_W(B)\ge0$ and
$2\kappa+\eta+2-\delta\le\rho-\kappa$, its cost is at most
\[
 \kappa(M_W+d_W)+(\rho-\kappa)(M_W-\cov_W(B))
 =\rho M_W+\kappa d_W-(\rho-\kappa)\cov_W(B).
\]
The cited procedure, weight modification, and restriction are polynomial time.
\end{proof}

\begin{corollary}[Path guarantee from Vygen's vertebrate-pair algorithm]
\label{cor:path-vygen-refined}
For every fixed $\varepsilon>0$, Vygen's vertebrate-pair algorithm,
the refined recursive reduction, and Lemma~\ref{lem:path-completion}
give a polynomial-time LP-relative asymmetric path TSP approximation with factor
\[
 7+\sqrt2+\sqrt{25+10\sqrt2}+\varepsilon.
\]
The path relaxation has integrality ratio at most
$7+\sqrt2+\sqrt{25+10\sqrt2}$.
\end{corollary}
\begin{proof}
Vygen~\cite[Theorem~3]{Vyg26} gives $\kappa=2$ and any
$\eta>4+2\sqrt2$. At the limiting value of $\eta$, equality
in~\eqref{eq:prior-refined-rho} gives the larger root
\[
 \rho=7+\sqrt2+\sqrt{25+10\sqrt2}.
\]
This is above $(15+\sqrt{73})/2$, so
Lemma~\ref{lem:path-completion} gives the same factor for paths. Rational parameters
sufficiently close to these limits satisfy the strict inequalities and
give the stated factor for every fixed $\varepsilon>0$. Taking the
infimum over $\varepsilon$ gives the integrality bound.
\end{proof}

This corollary matches the refined tour bound discussed by
Vygen~\cite[Section~4]{Vyg26}, improving his stated path factor
$15$~\cite[Section~5.2]{Vyg26}. It follows from the established
vertebrate-pair machinery and our transfer; the strengthened subtour cover
of Section~\ref{sec:cover} is used for the smaller common bound below.

\paragraph{Application to the new augmentation bound.}
The passage-dependent extension of our vertebrate-pair algorithm proves
the stronger recursive bound of Theorem~\ref{thm:augment}.
This additional bound is needed to reach the smaller common factor.
It satisfies~\eqref{eq:transfer-interface} with $\kappa=2$,
by the comparison immediately after Theorem~\ref{thm:augment}.
Its circulation normalization preserves positive support and unit laminar
boundary marginals, so it also preserves the component quantities and the
initial paths used in the transfer.
Since $R_*>(15+\sqrt{73})/2$, every rational $\rho>R_*$ chosen in
Section~\ref{subsec:parameter-choice} satisfies the condition for equal
factors in Lemma~\ref{lem:path-completion}. That lemma gives factor
$\rho$ for both tours and paths. This proves the polynomial-time
$(R_*+\varepsilon)$-approximation for every fixed $\varepsilon>0$ in
Theorem~\ref{thm:main}. Taking the infimum over $\varepsilon$ gives
integrality ratio at most $R_*$ for both relaxations.

\bibliographystyle{alphaurl}
\begingroup
\small
\raggedright
\bibliography{references}
\endgroup

\clearpage
\appendix
\section{Prescribed return paths and structured instances}
\label{sec:hub}
\label{app:hubs}

We prove stronger guarantees under explicit conditions on the boundary of each
laminar set. Throughout, the parameters of Svensson's algorithm are fixed,
their coefficient is $\alpha$, and the target tour ratio
$\rho\geq\max\{10,\alpha+2\}$ is rational with
$\theta\leq1-2/(\rho-2)$.
The child-specific prices below replace \eqref{eq:parameters} and
\eqref{eq:child-prices}, while the parameters of Svensson's algorithm
\eqref{eq:merging-parameters} are retained.

\subsection{Forcing a prescribed return}

Lemma~\ref{lem:witness} permits an arbitrary designated hitting set.
This is useful because the path to be forced is known before it belongs to
the integral output.

\begin{lemma}[A forced private visit]
\label{lem:hub-gate}
Suppose a private exterior vertex $r$ in the construction of
Theorem~\ref{thm:cover} has one incoming backward arc and one outgoing
forward arc, both of fractional value $m\in\{1/2,1\}$.
The cover can be required to visit $r$ exactly $2m$ times without changing
its other guarantees.  These visits survive in the terminal cover of
Theorem~\ref{thm:merge}.
\end{lemma}
\begin{proof}
Every witness assigns zero to the incoming arc and $m$ to the outgoing arc.
The upper copy of $r$ therefore has throughput $m$, transferred to its lower
copy by a downward switch.  Witness normalization preserves these values.
All cancellations and repairs take place in parts disjoint from the designated
vertex set, which contains $r$; they preserve its throughput even if another
endpoint of an incident arc is redirected.
Add an ordinary node gate at the upper copy of $r$, with both integral bounds
equal to $2m$.  The doubled comparison circulation remains feasible.
Integral minimum-cost circulation therefore retains the same cost guarantee
and forces exactly $2m$ visits.  Repairs preserve them, and Svensson's algorithm
retains every component meeting the designated set intact.
\end{proof}

For a proper region $W$, let $p^{\mathrm{in}}_u,p^{\mathrm{out}}_v$ be its unit entry and exit marginals.
Write $\operatorname{dist}_W$ for shortest-path distance in $G[W]$.
For a prescribed walk $P_{\mathrm{ret}}$ from $v^*$ to $u^*$ in $G[W]$, define
\begin{align*}
 \operatorname{disp}_W(v^*,u^*)
 &=\sum_v p^{\mathrm{out}}_v\operatorname{dist}_W(v,v^*)
   +\sum_u p^{\mathrm{in}}_u\operatorname{dist}_W(u^*,u),\\
 \operatorname{excess}_W(P_{\mathrm{ret}})&=c(P_{\mathrm{ret}})+\operatorname{disp}_W(v^*,u^*)-c_W^{\leftarrow}.
\end{align*}
The excess $\operatorname{excess}_W(P_{\mathrm{ret}})$ need not be nonnegative: $c_W^{\leftarrow}$ uses fixed nice
paths, whereas the connecting paths above are shortest paths.

\begin{lemma}[Prescribed return in the augmentation step]
\label{lem:hub-closure}
For a proper region $W$, at an induction step of
Theorem~\ref{thm:augment}, let $P_{\mathrm{ret}}$ meet the
inherited partial backbone $B$, or let $B=\varnothing$.  For
$m\in\{1/2,1\}$, the exterior construction can be modified to return an
Eulerian multiset $E_W\subseteq E[W]$ containing $2m$ copies of $P_{\mathrm{ret}}$,
such that $B\cup E_W$ is connected and spans $W$, with bound
\begin{equation}
 c(E_W)\leq
 \rho M_W+2e_W+2m\operatorname{excess}_W(P_{\mathrm{ret}})
       -(\rho-2)\cov_W(B\cup P_{\mathrm{ret}}).
 \label{eq:hub-augmentation}
\end{equation}
This modification uses the same descendant augmentation guarantees and the
same child inequalities as the original induction step.  In particular, it is
valid for the parameters of Theorem~\ref{thm:augment}.
\end{lemma}
\begin{proof}
If $M_W=0$, every internal arc has cost zero, and
$d_W=c_W^{\leftarrow}=b_W=e_W=\operatorname{disp}_W(v^*,u^*)=\operatorname{excess}_W(P_{\mathrm{ret}})=0$.
Close $P_{\mathrm{ret}}$ by any $u^*$--$v^*$ support path and include $2m$ copies of
this closed walk.  Add a path from $v^*$ to each vertex of $W$ and back.
Their union is a connected, spanning, zero-cost Eulerian multiset containing
the required copies of $P_{\mathrm{ret}}$, with every repetition retained.  If $B$ is
nonempty it intersects this multiset at a vertex of $P_{\mathrm{ret}}$.
Henceforth suppose $M_W>0$.

Let $Q$ be an optimal reverse coupling.  Keep its paired private returns with
values $(1-m)Q_{vu}$.  Add the following private two-arc passages:
$v\to o_v\to v^*$ of value $mp^{\mathrm{out}}_v$, with prescribed shortest lift;
$v^*\to h\to u^*$ of value $m$, with lift $P_{\mathrm{ret}}$; and
$u^*\to i_u\to u$ of value $mp^{\mathrm{in}}_u$, again with shortest lift.
Their assigned costs are the respective lift costs.
The resulting flow is balanced.  Set
\[
 B_0=(V(B)\cap W)\cup V(P_{\mathrm{ret}})\cup\{\text{private vertices}\},
\]
and contract the maximal proper laminar descendants missed by $B\cup P_{\mathrm{ret}}$.
Every real set avoiding $v^*,u^*$ receives exactly its original boundary inflow.
Consequently all parts and contracted children satisfy the cut and unit-degree
hypotheses of Theorem~\ref{thm:cover}.
The fractional cost before child contraction is
\[
 M_W-b_W+(1-m)c_W^{\leftarrow}
   +m\bigl(c(P_{\mathrm{ret}})+\operatorname{disp}_W(v^*,u^*)\bigr)
 =M_W+e_W+m\operatorname{excess}_W(P_{\mathrm{ret}}).
\]

Obtain a witness before splitting the exterior into private
vertices.  In the ordinary quotient with one exterior vertex $g$, the real
designated set $(V(B)\cap W)\cup V(P_{\mathrm{ret}})$ meets every remaining nonsingleton
laminar member.  Lemma~\ref{lem:witness} supplies a witness.
At least two real quotient vertices remain: $W$ has at least two vertices,
and the nonempty set $V(P_{\mathrm{ret}})$ is disjoint from every contracted proper
child.  Thus $W$ remains a nonsingleton structural set, making every private
rank smaller than every real rank.
Identifying all the new private vertices with $g$ changes this ordinary
circulation only by adding the two cycles
$g\to v^*\to g$ and $g\to u^*\to g$, each of value $m$.
Extend the witness by assigning their arcs out of $g$ to the lower layer and
their arcs into $g$ to the upper layer.  Any extra negative divergence occurs
only at the designated vertices $v^*,u^*$.
Now split $g$ according to the displayed private passages.  All private
incoming arcs are backward and all private outgoing arcs are forward, as in
Lemma~\ref{lem:private}.  Thus their upward switches remain zero.
Every retained segment between private visits still encounters a real
designated vertex.

Use Lemma~\ref{lem:hub-gate} at $h$.  The lifted terminal cover then contains
$2m\geq1$ copies of $P_{\mathrm{ret}}$.  Hence $B\cup P_{\mathrm{ret}}$ is an actual connected
subgraph after lifting, including when $B$ is empty.  It connects the real
designated vertices used in the formal identification. Every remaining
quotient intersection is actual or occurs at a contracted child, so the
component--child incidence graph is connected. Lemma~\ref{lem:joining}
therefore proves connectivity after the child completions.
The descendant calls concern sets missed by $B\cup P_{\mathrm{ret}}$.
In the cost calculation, the fractional comparison value changes by
$m\operatorname{excess}_W(P_{\mathrm{ret}})$ and the initial coverage becomes $\cov_W(B\cup P_{\mathrm{ret}})$.
Doubling the comparison cost proves \eqref{eq:hub-augmentation}.
\end{proof}

\subsection{A fixed-parameter criterion and a dispersion bound}

For each proper $L$ with $d_L>0$, fix a family
$\mathcal P_L$ of explicit walks of polynomial total encoding length,
containing all fixed nice returns
between its exit and entry ports. Define the completion credit, in the
original cost units, by
\begin{equation}
\begin{split}
 \mathrm{credit}_L=\min\Biggl\{d_L,\max\Biggl\{&
 \left(1-\frac2{\rho-2}\right)d_L,\\
 &\frac{(\rho-3)(c_L^{\leftarrow}+b_L)-(c_L^{\rightarrow}+b_L)}{\rho-2},
 \quad\max_{w\in L}2h_L(w),\\
 &
 \max_{\substack{P_{\mathrm{ret}}\in\mathcal P_L\\m\in\{1/2,1\}}}
 \left(\cov_L(P_{\mathrm{ret}})
       -\frac{2m}{\rho-2}\operatorname{excess}_L(P_{\mathrm{ret}})\right)
 \Biggr\}\Biggr\}.
\end{split}
 \label{eq:hub-credit}
\end{equation}
The four terms correspond to the maximum-coverage pair, the transportation
pair, a single-vertex partial backbone, and Lemma~\ref{lem:hub-closure} with empty
inherited partial backbone. The credit measures a cost saving in the completion
bound; it need not be the coverage of a particular backbone. Together with
reuse of a current component of coverage $\cov_L(B_L)$, these choices give the bound
\begin{equation}
 \rho M_L+2e_L-(\rho-2)\max\{\cov_L(B_L),\mathrm{credit}_L\}.
 \label{eq:hub-child-bound}
\end{equation}

For $0\leq\lambda\leq1$, define the following upper bound on the
remaining child charge:
\begin{align*}
 \operatorname{charge}_L(\lambda)
 ={}&2(c_L^{\leftarrow}+b_L)-2\lambda(c_L^{\rightarrow}+b_L)\\
 &-\bigl(\rho-\alpha-4-2\lambda\bigr)\mathrm{credit}_L\\
 &+\tfrac12[\alpha+6+4\lambda-\rho]_+
       (d_L-\mathrm{credit}_L).
\end{align*}

\begin{theorem}[Local sufficient criterion]
\label{thm:hub-criterion}
If every proper laminar member with $d_L>0$ satisfies
\begin{equation}
 \min_{0\leq\lambda\leq1}\operatorname{charge}_L(\lambda)\leq0,
 \label{eq:hub-criterion}
\end{equation}
then a polynomial-time algorithm returns a tour of cost at most
$\rho\LP$. The augmentation bound of Theorem~\ref{thm:augment} and
\eqref{eq:hub-augmentation} both hold with this $\rho$.
\end{theorem}
\begin{proof}
We prove the two augmentation statements simultaneously by induction on the
laminar family.  The case of zero descendant mass uses zero-cost paths as in
Theorem~\ref{thm:augment}.  For a positive child, choose a minimizing
$\lambda_L$ in \eqref{eq:hub-criterion} and set
\[
 \mathrm{price}_L=(1+\lambda_L)\mathrm{credit}_L
   +\tfrac14[\alpha+6+4\lambda_L-\rho]_+
       (d_L-\mathrm{credit}_L).
\]
In the comparison objective~\eqref{eq:global-cost}, charge
$\mathrm{price}_L/2$ at each endpoint of a child passage and use
$\lambda_L$ as the coefficient of the forward endpoint prices. Since
$\mathrm{credit}_L\geq(1-2/(\rho-2))d_L\geq3d_L/4$, we have
\[
 \mathrm{price}_L\geq(1+\lambda_L)\frac{3d_L}{4}
                    \geq\lambda_Ld_L.
\]
Thus real comparison costs remain nonnegative. Also
$\mathrm{price}_L\leq(1+\lambda_L)d_L$, because $\rho\geq\alpha+2$.

For actual component coverage $A\geq0$, after removing the common term
$\rho M_L$ and the boundary weight, the remaining child charge is at most
\begin{align*}
 2\mathrm{price}_L
 &+\alpha\max\{\theta d_L,\min\{A,d_L\}\}\\
 &+4[(1+\lambda_L)\min\{A,d_L\}-\mathrm{price}_L]_+\\
 &-2\lambda_L(c_L^{\rightarrow}+b_L)+2(c_L^{\leftarrow}+b_L)
       -(\rho-2)\max\{A,\mathrm{credit}_L\}.
\end{align*}
Here we used \eqref{eq:hub-child-bound} and
\[
 2e_L-2\lambda_Lc_L^{\rightarrow}
 \leq2(c_L^{\leftarrow}+b_L)-2\lambda_L(c_L^{\rightarrow}+b_L).
\]
The displayed expression is nondecreasing for
$0\leq A\leq\mathrm{credit}_L$, convex for
$\mathrm{credit}_L\leq A\leq d_L$, and decreasing for $A\geq d_L$.
Its value at $A=\mathrm{credit}_L$ is
$\operatorname{charge}_L(\lambda_L)$, and its value at $A=d_L$ is
\[
 \operatorname{charge}_L(\lambda_L)
 -[\rho-\alpha-6-4\lambda_L]_+(d_L-\mathrm{credit}_L).
\]
It is therefore nonpositive.  The boundary coefficient is
$2+\alpha\leq\rho$. These are precisely the child inequalities needed in
both induction steps, proving their bounds.  At the root, the
maximum-coverage initial pair of Lemma~\ref{lem:fresh} gives the tour bound.

All parameters and walk lists are rational and explicit. For fixed $\rho$,
$\operatorname{charge}_L(\lambda)$ is piecewise linear in $\lambda$; it suffices to
test $0,1$, and $(\rho-\alpha-6)/4$ when the latter lies in $[0,1]$.
The algorithm compares the finitely many completion scores and executes only
the selected descendant call.  Its running time is polynomial as in the
main construction.
\end{proof}

Define the boundary dispersion by
\[
 \operatorname{disp}(L)=\sum_{v,v'}p^{\mathrm{out}}_vp^{\mathrm{out}}_{v'}\operatorname{dist}_L(v,v')
          +\sum_{u,u'}p^{\mathrm{in}}_up^{\mathrm{in}}_{u'}\operatorname{dist}_L(u,u').
\]
\begin{corollary}[Concentrated boundary ports]
\label{cor:hub-dispersion}
Suppose $\operatorname{disp}(L)\leq\beta d_L$ for every proper laminar member with
$d_L>0$, where $0\leq\beta\leq2$. The tour guarantee is $\rho$ whenever
\[
 \rho\geq\alpha+6,
 \qquad(\rho-\alpha-6)(\rho-4)\geq2\beta.
\]
In the limit $\alpha\downarrow\alpha_*:=3+2\sqrt2$, the guarantee is
\[
 \frac{\alpha_*+10+
       \sqrt{(\alpha_*+2)^2+8\beta}}2.
\]
Every fixed additive relaxation is attained in polynomial time.  In particular,
$\beta=0$ gives limiting factor $\alpha_*+6$.
One may take the minimum of this bound and $R_*$ by also running the
unrestricted algorithm.
\end{corollary}
\begin{proof}
Draw $(v^*,u^*)$ from an optimal reverse coupling and choose
$P_{\mathrm{ret}}=P^L_{v^*u^*}$ with $m=1/2$.
Lemma~\ref{lem:coverage} gives expected coverage $c_L^{\leftarrow}+b_L$;
the expected walk cost is $c_L^{\leftarrow}$ and the expected dispersion is $\operatorname{disp}(L)$.
Consequently \eqref{eq:hub-credit} gives
\[
 \mathrm{credit}_L\geq
 \max\left\{\left(1-\frac2{\rho-2}\right)d_L,
             c_L^{\leftarrow}+b_L-\frac{\operatorname{disp}(L)}{\rho-2}\right\}.
\]
Indeed, the maximum of the prescribed-return credits is at least their
expectation, and clipping at $d_L$ preserves this lower bound because
$c_L^{\leftarrow}+b_L\leq d_L$. With $\lambda=0$ and $\rho\geq\alpha+6$, the
criterion reduces to
$(\rho-\alpha-4)\mathrm{credit}_L\geq2(c_L^{\leftarrow}+b_L)$.
As a function of $c_L^{\leftarrow}+b_L\in[0,d_L]$, the expression
\[
 (\rho-\alpha-4)
 \max\left\{\left(1-\frac2{\rho-2}\right)d_L,
             c_L^{\leftarrow}+b_L-\frac{\beta d_L}{\rho-2}\right\}
 -2(c_L^{\leftarrow}+b_L)
\]
is minimized where the two branches meet, at
$c_L^{\leftarrow}+b_L=(1-(2-\beta)/(\rho-2))d_L$. Its value there is
\[
 \frac{d_L}{\rho-2}\bigl((\rho-\alpha-6)(\rho-4)-2\beta\bigr)\geq0.
\]
Thus the criterion holds.
The limiting expression solves its quadratic equality.

Finally, all fixed nice path costs are at most $d_L$ and distances on the
diagonal are zero.  Thus the useful purely marginal sufficient bound is
\[
 \operatorname{disp}(L)\leq
 \left(2-\sum_v (p^{\mathrm{out}}_v)^2-\sum_u (p^{\mathrm{in}}_u)^2\right)d_L\leq2d_L.
\]
\end{proof}

\subsection{Replacing a passage when the ports are unique}

\begin{lemma}[Passage replacement]
\label{lem:hub-replacement}
Suppose the descendant augmentation guarantees and child inequalities of
Lemma~\ref{lem:hub-closure} hold for a child $L$.
Let a lifted Eulerian parent multiset contain a charged passage
$P_{\mathrm{old}}$ from $u$ to $v$ inside $L$.
For a walk $P_{\mathrm{ret}}$ from $v$ to $u$, that child can be
completed while replacing $P_{\mathrm{old}}$, with net cost increase at most
\[
 \begin{aligned}
 \rho M_L+c(P_{\mathrm{ret}})-c(P_{\mathrm{old}})
   &+2\operatorname{disp}_L(v,u)-2b_L\\
   &-(\rho-2)\cov_L(P_{\mathrm{ret}}).
 \end{aligned}
\]
The resulting multiset is Eulerian and joins every current component meeting
$L$.
\end{lemma}
\begin{proof}
Apply Lemma~\ref{lem:hub-closure} with empty backbone and $m=1$.
Its connected spanning output $E_L$ contains two copies of $P_{\mathrm{ret}}$ and costs
at most
$\rho M_L+2c(P_{\mathrm{ret}})+2\operatorname{disp}_L(v,u)-2b_L-(\rho-2)\cov_L(P_{\mathrm{ret}})$.
Delete $P_{\mathrm{old}}$ from the parent and one copy of $P_{\mathrm{ret}}$ from $E_L$.
The two removed paths have opposite divergences, so their deletion preserves
balance in the union.  Deleting one of two copies of $P_{\mathrm{ret}}$ preserves the
entire underlying arc support of $E_L$, and hence its connectivity and
spanning property.  It contains both endpoints of $P_{\mathrm{old}}$, reconnects any
pieces separated by its deletion, and meets every other component incident
with $L$.  Subtracting the two removed costs proves the bound.
\end{proof}

\begin{theorem}[One entry and one exit]
\label{thm:hub-single-ports}
Suppose every proper laminar member $L$ with $d_L>0$ has one entry vertex and
one exit vertex in the positive support.  A polynomial-time algorithm has tour
guarantee $\rho$ whenever
\[
 \rho\geq\alpha+6-\frac4{\rho-1}.
\]
The limiting value is $\rho_1$, the larger root of
$\rho_1^2-(\alpha_*+7)\rho_1+\alpha_*+10=0$.
\end{theorem}
\begin{proof}
Neither unique port lies in a proper positive descendant $J$ of $L$.
Indeed, positive support and strong connectivity give
$x(L\setminus J\to J)>0$, so tightness of $J$ implies that the exterior
entry mass at $J$ is strictly less than one; the exit statement is symmetric.
Thus $b_L=0$.

Prove the augmentation statements simultaneously as above, but at each such
child use $\lambda=0$ and charge $c_L^{\rightarrow}/2$ at each endpoint of a passage.
Every charged passage is the same nice entry--exit path, of coverage
$c_L^{\rightarrow}$, so its total price is exact and no lifting correction is needed.
Apart from $\rho M_L$ and the boundary weight, the total child charge is at most
\[
 \begin{aligned}
 2c_L^{\rightarrow}+\alpha\max\{\theta d_L,c_L^{\rightarrow}\}
 +\min\Bigl\{&2c_L^{\leftarrow}-(\rho-2)
   \max\Bigl\{c_L^{\rightarrow},\Bigl(1-\frac2{\rho-2}\Bigr)d_L\Bigr\},\\
 &-c_L^{\rightarrow}-(\rho-3)c_L^{\leftarrow}\Bigr\}.
 \end{aligned}
\]
The first completion choice uses reuse or the maximum-coverage pair.
The second uses Lemma~\ref{lem:hub-replacement} with the nice reverse
path: its dispersion is zero and its coverage equals its cost $c_L^{\leftarrow}$.
For $c_L^{\rightarrow}\geq(1-2/(\rho-2))d_L$, the two affine functions of $c_L^{\leftarrow}$ cross at
$c_L^{\leftarrow}=(\rho-3)c_L^{\rightarrow}/(\rho-1)$; their lower envelope is at most its value there.
The displayed charge is consequently at most
\[
 c_L^{\rightarrow}\left(\alpha+6-\rho-\frac4{\rho-1}\right)\leq0.
\]
For $c_L^{\rightarrow}<(1-2/(\rho-2))d_L$, distribute the preceding parent charge into the two
branches. For fixed $c_L^{\leftarrow}$ and $d_L$, both branches are nondecreasing in
$c_L^{\rightarrow}$. Their minimum is therefore bounded by its value at
$c_L^{\rightarrow}=(1-2/(\rho-2))d_L$, already handled.
Boundary weights are covered because $\rho\geq\alpha+2$.
The same induction and root initialization finish the proof; solving the
limiting equality gives the stated quadratic.
\end{proof}

\begin{remark}[Path guarantees]
The dispersion bound transfers to path TSP with factor $\rho$ when every
positive laminar member of the normalized path instance satisfies its
condition, since $\rho\ge\alpha_*+6>(15+\sqrt{73})/2$. The temporary SCC
regions require the condition only on their proper laminar descendants.
The one-entry/one-exit factor $\rho_1$ lies below the threshold for a common
tour and path factor at $\kappa=2$ in Lemma~\ref{lem:path-completion}.
\end{remark}

\end{document}